\documentclass{aa}

\usepackage{graphicx}
\usepackage{txfonts}
\usepackage[version=4]{mhchem}
\usepackage{makecell}

\begin{document}

   \title{Abundances of trace constituents in Jupiter's atmosphere inferred from Herschel/PACS\thanks{{\it Herschel} is an ESA space observatory with science instruments provided by European-led Principal Investigator consortia and with important participation from NASA.} observations}

   \author{C. Gapp\inst{1,2}\fnmsep\thanks{\email{gapp@mpia.de}}\orcid{0009-0007-9356-8576}
          \and M. Rengel\inst{2}\orcid{0000-0001-6419-4893}
          \and P. Hartogh\inst{2}\orcid{0000-0002-9550-6551}
          \and H. Sagawa\inst{3}
          \and H. Feuchtgruber\inst{4}
          \and E. Lellouch\inst{5}
          \and G. L. Villanueva\inst{6}\orcid{0000-0002-2662-5776}
          }

   \institute{Max-Planck-Institut für Astronomie, 69117 Heidelberg, Germany
         \and Max-Planck-Institut für Sonnensystemforschung, 37077 Göttingen, Germany
         \and Kyoto Sangyo University, Kyoto 603-8555, Japan
         \and Max-Planck-Institut für Extraterrestrische Physik, 85746 Garching, Germany
         \and LESIA–Observatoire de Paris, CNRS, UPMC Univ. Paris 06, Univ. Denis Diderot, Sorbonne Paris Cite, Meudon, France
         \and NASA Goddard Space Flight Center, Greenbelt, MD, USA
             }

   \date{Received ; accepted }
 
  \abstract
   {On October 31, 2009, the Photodetector Array Camera and Spectrometer (PACS) on board the \textit{Herschel} Space Observatory observed far-infrared spectra of Jupiter in the wavelength range between 50 and 220$\,\mu$m as part of the program "Water and Related Chemistry in the Solar System". The spectra have an effective spectral resolution between 900 and 3500, depending on the wavelength and grating order.}
   {We investigate the disk-averaged chemical composition of Jupiter's atmosphere as a function of height using these observations.}
   {We used the Planetary Spectrum Generator (PSG) and the least-squares fitting technique to infer the abundances of trace constituents.}
   {The PACS data include numerous spectral lines attributable to ammonia (NH$_3$), methane (CH$_4$), phosphine (PH$_3$), water (H$_2$O), and deuterated hydrogen (HD) in the Jovian atmosphere and probe the chemical composition from $p\sim 275$\,mbar to $p\sim 900$\,mbar. From the observations, we infer an ammonia abundance profile that decreases from a mole fraction of $(1.7\pm 0.8)\times 10^{-4}$ at $p\sim 900$\,mbar to $(1.7\pm 0.9)\times 10^{-8}$ at $p\sim 275$\,mbar, following a fractional scale height of about $0.114$. For phosphine, we find a mole fraction of $(7.2\pm 1.2)\times 10^{-7}$ at pressures higher than $(550\pm 100)$\,mbar and a decrease of its abundance at lower pressures following a fractional scale height of $(0.09\pm 0.02)$. Our analysis delivers a methane mole fraction of $(1.49\pm 0.09)\times 10^{-3}$. Analyzing the HD $R(0)$ line at $112.1\,\mu$m yields a new measurement of Jupiter's D/H ratio, $\text{D/H}=(1.5\pm 0.6)\times 10^{-5}$. Finally, the PACS data allow us to put the most stringent $3\sigma$ upper limits yet on the mole fractions of hydrogen halides in the Jovian troposphere. These new upper limits are $<1.1\times 10^{-11}$ for hydrogen fluoride (HF), $<6.0\times 10^{-11}$ for hydrogen chloride (HCl), $<2.3\times 10^{-10}$ for hydrogen bromide (HBr) and $<1.2\times 10^{-9}$ for hydrogen iodide (HI) and support the proposed condensation of hydrogen halides into ammonium halide salts in the Jovian troposphere.}
   {}


   \maketitle

\section{Introduction}
The chemical composition of Jupiter's atmosphere yields insight into the history of the Solar System and the interaction of Jupiter with other Solar System bodies such as comets. Additionally, Jupiter is an important benchmark for giant planet atmospheres that both applies to other Solar System planets and exoplanets (see, e.g., \citealp{heng21}). Thus, it has been studied extensively using both ground- and space-based observations, laboratory studies, theoretical models, and one in situ survey carried out using the \textit{Galileo} Entry Probe (\citealp{zahn96,mahaffy98,niemann98,wong04}). Observations of Jupiter's infrared spectrum have been especially useful for investigations of the chemical and thermal structure of the Jovian atmosphere as well as its spatial and temporal variability thanks to the intensity of Jupiter's thermal radiation and the wealth of molecular spectral lines in the infrared. Therefore, the Jovian spectrum has been observed frequently from near-infrared (NIR) wavelengths (see, e.g., \citealp{drossart82,kunde82,bjoraker86,carlson93,encrenaz96,noll96,irwin98,fouchet00nh3,fouchet00hydro,lellouch02,roos04,bjoraker15,grassi17}), where the Jovian thermal radiation starts to dominate over reflected Solar radiation, to mid-infrared (MIR) wavelengths (see, e.g., \citealp{gautier81,gautier82,knacke82,kunde82,carlson93,encrenaz96,fouchet00nh3,fouchet00hydro,lellouch01,lellouch02,achterberg06,nixon07,nixon10,fletcher09,fletcher16,pierel17}). However, there are remarkably few studies that targeted the far-infrared (FIR) range in Jupiter (see, e.g., \citealp{burgdorf02,fouchet04,cavalie13,teanby14,pierel17}), even though it bridges the gap between MIR and submillimeter wavelengths that have also been used to study Jupiter's atmosphere (see, e.g., \citealp{marten95,lellouch02,moreno03,cavalie08,dePater19_ALMA,cavalie21}).

In this study, we present and analyze observations of Jupiter carried out with the Photodetector Array Camera and Spectrometer (PACS, \citealp{pacs10}) on board the \textit{Herschel} Space Observatory (\textit{Herschel}, \citealp{herschel1}) in the FIR at wavelengths between 50 and 220$\,\mu$m that contribute to a more complete understanding of the Jovian spectrum. We report the gases identified in the spectra, constrain their abundances, and derive upper limits for the mole fractions of hydrogen halides that remain undetected.

\section{Observations and data reduction} \label{sec:jupiter_observations}
In the framework of the key program "Water and Related Chemistry in the Solar System" \citep{hsso}, \textit{Herschel} observed Jupiter employing PACS as a spectrometer on October 31, 2009. Two separate observations with observation identifiers (ObsID) 1342186573 and 1342186574 were conducted to build up the full wavelength range between 50 and 220$\,\mu$m. The observational details of both observations and selected corresponding ephemerides (taken from the NASA/JPL Horizons Web-Interface\footnote{\url{https://ssd.jpl.nasa.gov/horizons.cgi}}) are listed in Table \ref{tab:observations}.

During both observations, PACS obtained spectra using its array of $5\times 5$ spatial image pixels (spaxels) with beams with full-widths at half maximum (FWHMs) of approximately $9''$ each. The integral field unit covered a total field of view (FOV) of $47''\times 47''$ on the sky. The spectral range of the blue spectrometer was covered in grating order 3 for the first observation (ObsID 1342186573) and in grating order 2 for the second observation (ObsID 1342186574). During both observations, the spectra of the red spectrometer were covered using grating order 1. We denominate the corresponding spectral ranges using B2 and B3 for the blue spectrometer in the grating orders 2 and 3, respectively, and R1 for the red spectrometer in grating order 1. The  resulting data have not been released to the \textit{Herschel} Science Archive because they required specialized efforts on the calibration (see Subsec. \ref{subsec:data_reduction}), but are available upon request.
\begin{table*}[htbp!]
	\caption{PACS observations of Jupiter analyzed in this study with selected ephemerides.}
	\label{tab:observations}
	\centering
	\begin{tabular}{c c c}
		\hline\hline
		\textbf{Parameter / ObsID} & \textbf{1342186573} & \textbf{1342186574} \\ \hline
		Operation day & 170 & 170 \\
		Start time [UTC] & 10/31/2009 15:59:55 & 10/31/2009 18:12:35 \\
		Duration [s] & 7825 & 7825 \\
		Blue section grating order and wavelength range & B3: $50-73\,\mu$m & B2: $68-105\,\mu$m \\
		Red section grating order and wavelength range [$\mu$m] & R1: $101-220\,\mu$m & R1: $101-220\,\mu$m \\
		\textit{Herschel}'s Sub-Jovian latitude [$^\circ$] & 0.31 & 0.31 \\
		\textit{Herschel}'s Sub-Jovian longitude [$^\circ$] & 161.3 & 241.7 \\
		Distance \textit{Herschel}-Jupiter [AU] & 4.76 & 4.77 \\
		Relative velocity between \textit{Herschel} and Jupiter [km\,s$^{-1}$] & 28.16 & 28.18 \\
		Jupiter's angular diameter ["] & 41.38 & 41.37 \\ \hline
	\end{tabular}
    \tablefoot{All ephemerides given here are the respective values' minute-wise averages during each observation.}
\end{table*}

\subsection{Applied observation template} \label{subsec:aot}
During both observations, the PACS measurements were carried out using the unchopped full resolution up-and-down grating scan calibration observation template. Standard chopped-and-nodded astronomical observation templates could not be used since the flux of Jupiter on the PACS detectors was so large that persistence effects due to on-and-off transients would have dominated the measured signal, preventing any quantitative analysis. Furthermore, the large signal of Jupiter on the detectors required the integration time to be shortened to prevent saturation. As a consequence, more data had to be transmitted to the onboard storage per unit time, exceeding the nominal bandwidth allocation. For this purpose, the time-limited special telemetry burst mode had to be activated during these two observations.

\subsection{Data reduction} \label{subsec:data_reduction}
The data processing from Level 0 to 1 was performed using the standard PACS pipeline modules of the \textit{Herschel} interactive processing environment (HIPE V6). All subsequent processing was carried out with standard interactive data language (IDL) tools. The main distortion of these data at Level 1 was caused by cosmic ray hits penetrating into the detectors. These impacts caused response transients which decayed on time scales of few to tens of seconds. Strong transients have been masked out entirely before further processing. At the beginning of the observations, the data were somewhat affected by detector drifts caused by the transient from low signal to the large signal increase when pointing toward Jupiter. Therefore, the scan of the grating from long to short wavelengths was considered more stable compared to the initial scan upward in wavelength. The data of all scans and all 16 pixels per spaxel were therefore scaled by low order polynomials to the mean spectral shape of the down scan. In order to remove the remnants of smaller cosmic ray hits, iterative sigma clipping was then applied. The resulting spectra were then rebinned to appropriate resolutions depending on the wavelength range.

It would have been desirable to analyze the data spaxel-wise to infer information on the latitudinal and longitudinal variability of the Jovian atmosphere. Unfortunately, the high fluxes on the spaxel array and the subsequently unavoidable use of the unchopped observation template (see Subsec. \ref{subsec:aot}) compromised the signal-to-noise ratio (S/N) of each spaxel substantially. Thus, to increase the S/N, we only analyzed the average of the innermost nine spaxels, yielding information on the disk-averaged properties of Jupiter's atmosphere. We increased the S/N of the data of the R1 spectral band further by averaging the two measurements and only analyzing the resulting average. Since the absolute calibration accuracy is limited by detector response drifts and slight pointing offsets, the resulting spectra were divided by their local continua for the following analysis.

\subsection{Effective spectral resolution} \label{subsec:effective_resolution}
PACS's spectral resolution for point sources depends on the grating order of the measurements as well as the wavelengths and varies between 990 and 5500 \citep{pacs_observers_manual}. The effective spectral resolution ($r_{eff}$) of the Jupiter spectra deviates from that spectral resolution due to Jupiter's extended nature that causes smearing between wavelengths in the spectra as well as, to a lesser extend, the planet's rotation\footnote{With an equatorial rotational speed of $v\sim 10$\,km\,s$^{-1}$, Jupiter's rotation causes rotational broadening on the order of $R=c/v\sim 30,000$, thus on a much higher resolution than the PACS spectra.}. We determined the effective spectral resolution of the Jupiter spectra as a function of the wavelength ($\lambda$) and the lowest resolvable difference in wavelength ($\Delta\lambda_{eff}$) using
\begin{equation}
	r_{eff}(\lambda) = \frac{\lambda}{\Delta \lambda_{eff}}. \label{eq:R}
\end{equation}

\begin{figure}[htbp!]
	\centering
	\includegraphics[width=\hsize]{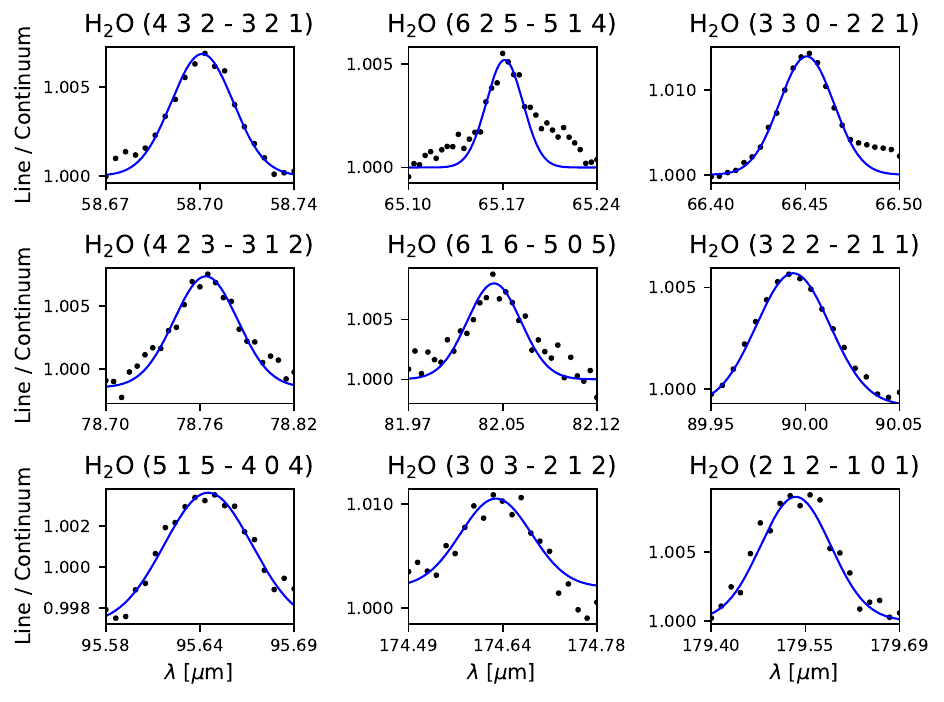}
	\caption{Confirmed emission lines caused by the presence of water vapor in the upper Jovian stratosphere. The PACS measurements are shown with black dots and the Gaussians fit to the data to obtain the FWHM of the lines are plotted using blue lines. The rotational transitions that cause the lines are given in the panels' titles using the three rotational quantum numbers of the upper and lower states, respectively.}
	\label{fig:H20_lines}
\end{figure}

Emission lines of water caused by Jupiter's stratospheric water reservoir are well suited to determine the effective spectral resolution of the PACS spectra, mainly for two reasons: Firstly, water emits numerous spectral lines in PACS's wavelength range. And secondly, as a molecule mostly residing in Jupiter's upper stratosphere (see, e.g., \citealp{lellouch02,cavalie08,cavalie13}), its emission lines experience little collisional and Doppler broadening due to the low pressures and moderate temperatures there. The reason for water vapor to be much less abundant in the lower stratosphere is that it freezes out at its low temperature. Therefore, the water vapor's lines' widths are very small and the exact lineshapes cannot be resolved with PACS's spectral resolution. So, all broadening of the water lines in the observed spectra can be attributed to the finite spectral resolution of PACS \citep{pacs08} and hence the shapes of the lines do not hold any information about the abundance profile of water in Jupiter's stratosphere. Hence, we used the FWHM of the lines as $\Delta\lambda_{eff}$ in Equation \ref{eq:R} instead of analyzing the Jovian stratospheric water reservoir. To obtain the FWHM of the lines, we fit Gaussians to the data and then calculated the Gaussians' FWHM. In total, we detected nine water emission lines (three lines from the B3 spectral band, four from the B2 spectral band and two from the R1 spectral band) and used all of them for the determination of the effective spectral resolution. These water emission lines and the respectively fit Gaussians are shown in Fig. \ref{fig:H20_lines}.

\begin{figure}[htbp!]
	\centering
	\includegraphics[width=\hsize]{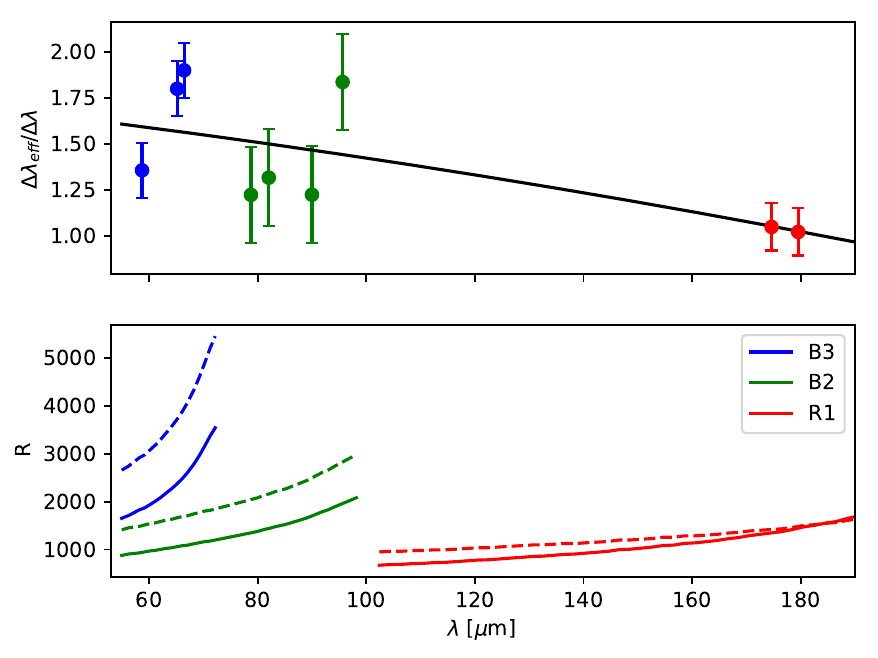}
	\caption{Determination of PACS's effective spectral resolution $R$ over its spectral range using water emission lines. \textit{Upper panel:} The colored dots show the ratios of the effective resolvable difference in wavelength and the resolvable difference in wavelength given by PACS's spectral resolution for point sources determined from the detected water emission lines. The dots are color-coded by PACS's spectral band they fall into. The solid black line shows a second-order polynomial fit to the data. \textit{Lower panel:} PACS's resolution for point sources in each spectral band, as given in the observer's manual \citep{pacs_observers_manual} is shown using dashed lines, and the effective spectral resolution of the Jupiter observations as determined from the polynomial fit is plotted using solid lines.}
	\label{fig:R}
\end{figure}

We then used the determined FWHM to calculate the ratios of the determined $\Delta\lambda_{eff}$ to the $\Delta\lambda$ given in the PACS observer's manual for point sources, fit a polynomial of second order to these ratios and calculated $r_{eff}$ of every grating order using Equation \ref{eq:R} and the polynomial fit (see Fig. \ref{fig:R}). Both confirmed lines in the R1 spectral band are consistent with $\Delta\lambda_{eff}/\Delta\lambda=1$. So, the spectral resolution for point sources appears to be a good approximation of the effective spectral resolution in that spectral band and thus, we used the spectral resolution for point sources as $r_{eff}$ in all following analyses of the R1 spectral band. For the B3 and B2 bands, we used the effective spectral resolutions derived from the water emission lines for all subsequent analyses.

\subsection{Spectral leakage and ghosts}
It is a known issue of the PACS data that radiation from one grating order can leak into another grating order due to the finite steepness of the order sorting filter cut-off edges. None of the regions known to include leakage effects (see \citealp{pacs_observers_manual}) are analyzed in this study. Another known issue are spectral ghosts which are strong emission lines from one PACS spaxel that show up with lower fluxes at a different wavelength in another spaxel. We reviewed all data that could contain these ghosts by inspecting both the data around the wavelengths from which ghosts could originate and the data around the wavelengths at which the ghost lines would occur and did not identify any spectral ghosts. 

\subsection{Data overview} \label{subsec:data_overview}
\begin{figure*}[htbp!]
    \centering
    \includegraphics[width=\hsize]{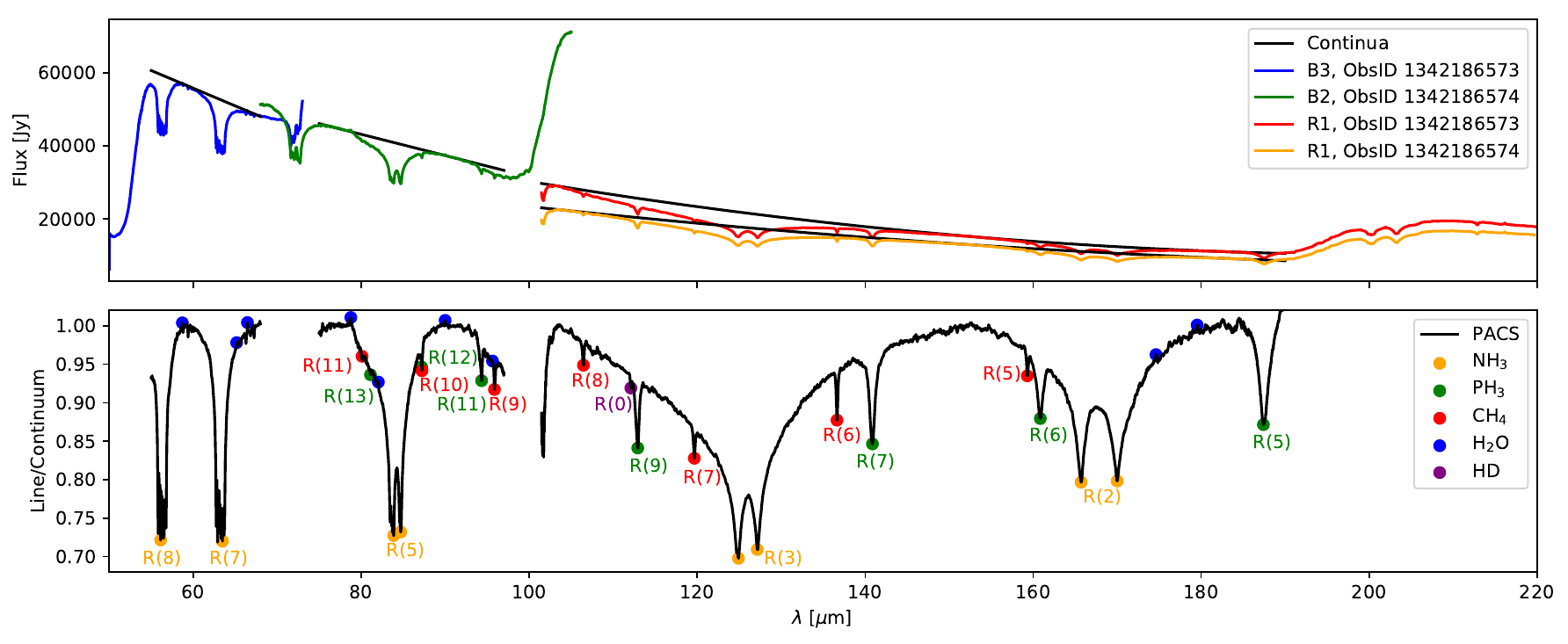}
    \caption{PACS spectra of Jupiter, averaged over the nine innermost spaxels. \textit{Upper panel:} The absolute fluxes are shown using colored solid lines and the polynomials used to approximate the continua of each spectral band are depicted using black solid lines. The polynomials' coefficients are listed in Table \ref{tab:continua}. \textit{Lower panel:} The observations expressed in line-to-continuum ratios are plotted using black solid lines with the identified spectral features indicated with colored dots and the respective molecule's rotational line transition. The parts of the data that are affected by spectral leakage are not shown in this panel. The data at wavelengths longer than $100\,\mu$m are the average of the two measurements in the R1 spectral band after normalization by the respective continua.}
    \label{fig:inventory}
\end{figure*}

The PACS observations, before and after normalization by the continua, are plotted in Fig. \ref{fig:inventory}. However, the absolute fluxes presented in the upper panel of that figure are likely inaccurate due to the named challenges of the observations (see Subsec. \ref{subsec:data_reduction}). Thus, we only analyzed the line-to-continuum ratios. To approximate the continua for the normalization, we fit polynomials of the shape
\begin{equation}
    f(\lambda) = a_0 + a_1\lambda + a_2\lambda^2 \label{eq:poly}
\end{equation}
to selected parts of the data far from any absorption or emission features. All wavelength ranges used to fit polynomials for approximating the continua and the resulting polynomial coefficients ($c_0$, $c_1$ and $c_2$) are listed in Table \ref{tab:continua}.
\begin{table*}[htbp!]
	\caption{Polynomial fits used to approximate the continuum of each spectral band and observation.}
	\label{tab:continua}
	\centering
	\begin{tabular}{c c c c c c}
		\hline\hline
		\textbf{Spectral band} & \textbf{ObsID} & \textbf{Wavelengths used for the polynomial fit [$\mu$m]} & $a_0$ [Jy] & $a_1$ [Jy\,$\mu$m$^{-1}$] & $a_2$ [Jy\,$\mu$m$^{-2}$] \\ \hline
		B3 & 1342186573 & $58.95-59.25$, $67.4-67.7$ & 113815 & -967.280 & \\
		B2 & 1342186574 & $76.7-77.0$, $91.6-91.9$ & 89631.3 & -580.734 & \\
		R1 & 1342186573 & $103.4-103.7$, $152.0-152.3$, $181.7-182.0$ & 83536.0 & -743.613 & 1.80825 \\
		R1 & 1342186574 & $103.4-103.7$, $152.0-152.3$, $181.7-182.0$ & 57321.8 & -430.794 & 0.916970 \\ \hline
	\end{tabular}
    \tablefoot{Equation \ref{eq:poly} gives the mathematical expression to calculate each polynomial using its coefficients listed here. We set $a_2=0$ for the approximation of the continua in the spectral bands B3 and B2.}
\end{table*}

We labeled the observed spectral features using $R(J'')$, where $J''$ is the rotational quantum number of the lower energy state of the rotational transition that creates the spectral line. We use this labeling scheme for the rest of this work, also stating the absorbing molecule before each transition label. The only exception from this labeling scheme are the water emission lines (Fig. \ref{fig:H20_lines}) which we identify using the whole set of three rotational quantum numbers of the upper and lower states, respectively.

\section{Radiative transfer modeling} \label{sec:model_setup}
We computed the emerging radiance using a forward model embedded in the online radiative transfer tool Planetary Spectrum Generator\footnote{\url{https://psg.gsfc.nasa.gov/}} (PSG, \citealp{psg}). PSG is a highly sophisticated, versatile and computationally effective radiative transfer tool that calculates model spectra by employing the standard equations of radiative transfer and applying spectral line data from nine different spectral line catalogs.

\begin{figure}[htbp!]
	\centering
	\includegraphics[width=\hsize]{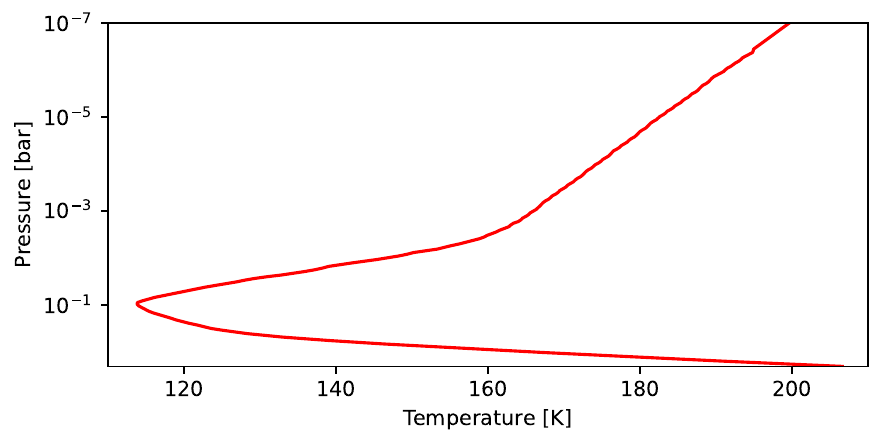}
	\caption{Temperature profile \citep{seiff98_galileo,nixon07} adopted for this analysis.}
	\label{fig:pt}
\end{figure}

We used HITRAN \citep{hitran} for both rotational lines and collision-induced absorption and employed ephemerides taken from the NASA/JPL Horizons Web-Interface for the simulations. We adopted the a priori temperature profile that \cite{nixon07} derived from the temperature measurements of the \textit{Galileo} Entry Probe \citep{seiff98_galileo} and was also later used as a reference temperature profile by \cite{fletcher09}. The adopted temperature profile is plotted in Fig. \ref{fig:pt}. The abundances of molecular hydrogen and helium ($86.2\,\%$ and $13.6\,\%$, respectively) were taken from the results of the \textit{Galileo} entry probe \citep{niemann98,zahn96}.

To compare the simulated spectra to the line-to-continuum ratios of the PACS data, we also normalized the simulated spectra to line-to-continuum ratios. We used the same wavelength ranges as for the PACS data's normalization as input for polynomial fits to approximate the model spectra's continua.

\section{Statistical methods}
To determine the mole fractions of the observed molecules in Jupiter's atmosphere with PACS, their uncertainties as well as upper limits for undetected species, we simultaneously applied the least-squares method to the synthetic and measured spectra and calculated the best fit by minimizing
\begin{equation}
	\chi^2 = \sum_{i=1}^N \frac{(D_i-M_i(\vec{a}))^2}{\sigma_{D,i}^2 + \sigma_{M,i}^2}. \label{eq:chi-squared}
\end{equation}
Here, $D_i$ are the PACS data points, $M_i$ are the model predictions for each data point generated employing a set of model parameters $\vec{a}$, $\sigma_{D,i}$ are the uncertainties of the PACS data, $\sigma_{M,i}$ are the uncertainties of the model predictions and $N$ is the total number of data points. We found that $2\,\%$ was appropriate for all $\sigma_{D,i}$ and approximated $\sigma_{M,i}$ by averaging the deviation between the data and the best-fitting model outside of the spectral lines.

\subsection{Detections and upper limits} \label{subsec:upper-limits}
To determine, whether or not a molecule is significantly detectable using the PACS data, we calculated the difference in $\chi^2$ between model spectra assuming the molecule at different vertically constant mole fractions in the Jovian atmosphere and a model spectrum excluding the target molecule. This translates to
\begin{equation}
    \Delta\chi_0^2(a) = \chi^2(a) - \chi^2(0), \label{eq:delta_chi-squared0}
\end{equation}
where $\chi^2(0)$ is the $\chi^2$ value of the model without the molecule and $\chi^2(a)$ refers to the models including the molecule with a mole fraction $a$. When $a_{min}$ is the mole fraction that minimizes $\chi^2$, then $\Delta\chi^2_0(a_{min})$ has to be lesser than $-1$, $-4$ or $-9$ for a detection at the $1$, $2$ or $3\sigma$ significance levels, respectively \citep{nr}. When the detection of a molecule was not possible at the $2\sigma$ level at least, we instead determined its $3\sigma$ upper limit.

\subsection{Confidence intervals}
To set $1\,\sigma$ confidence intervals on the parameters that determine each detected molecule's abundance profile, we calculated
\begin{equation}
    \Delta\chi^2(\vec{a}) = \chi^2(\vec{a}) - \chi_{min}^2(\vec{a}_{best}) \label{eq:delta_chi-squared}
\end{equation}
for a grid of values of the abundance profile parameters $\vec{a}$. Here, $\chi^2(\vec{a})$ is the $\chi^2$ value between the PACS data and the model spectrum generated using a set of values for the model parameters $\vec{a}$ and $\vec{a}_{best}$ are the values of the abundance profile parameters that minimize $\chi^2$. To explore an adequate parameter grid that incorporates the $1\,\sigma$ ranges of all parameters, we first ran the simulations on wide parameter grids and then reduced the grid size around the best fits of the initial grid searches.

To determine the parameters' $1\,\sigma$ intervals, we drew contours with the $\Delta\chi^2$ value that correspond to the $1\,\sigma$ interval for the appropriate number of adjustable abundance profile parameters. We took the maximum and minimum values of this contour as the limits of the $1\,\sigma$ confidence interval for each parameter (see, e.g., \citealp{nr}).

\section{Results: Abundances of trace constituents}
We analyzed the PACS data by fitting the molecular lines molecule-by-molecule, adding one analyzed molecule to the model of the chemical composition of Jupiter's atmosphere after the other. The molecule that adds the most opacity to the Jovian atmosphere in the observed wavelength range is ammonia, so we first analyzed ammonia and then added analyses of phosphine, methane, deuterated hydrogen and hydrogen halides.

\subsection{Ammonia} \label{subsec:nh3}
Ammonia plays a vital role for the visual appearance of Jupiter, since it condenses into two kinds of clouds at different pressure levels: firstly, the uppermost cloud layer of Jupiter which is composed of ammonia ice and secondly, the second uppermost cloud layer which is made of ammonium hydrosulfide that results from the reaction of ammonia and hydrogen sulfide (see, e.g., \citealp{lewis69,weidenschilling73,atreya85,encrenaz96,irwin98,galileo99clouds}). Additionally, ammonia is subject to photodissociation in Jupiter's upper troposphere, further depleting that part of the atmosphere from it. We parameterized its vertical abundance profile using a three-parameter model of the shape
\begin{equation}
	a_{\ce{NH_3}}(p) =
	\begin{cases}
		a_{\ce{NH_3}}(\infty) & \text{if } p\ge p_{\ce{NH_3}} \\
		a_{\ce{NH_3}}(\infty)\left(\frac{p}{p_{\ce{NH_3}}}\right)^{((1-f_{\ce{NH_3}})/f_{\ce{NH_3}})} & \text{else} \label{eq:nh3_profile}
	\end{cases}
\end{equation}
(see, e.g., \citealp{pierel17}) for the following analysis. Here, $a_{\ce{NH_3}}(\infty)$ is the vertically constant ammonia mole fraction below the knee pressure $p_{\ce{NH_3}}$ above which the ammonia abundance starts to decrease. The parameter $f_{\ce{NH_3}}$ is ammonia's fractional scale height that governs the decrease of the ammonia abundance above the knee pressure level.

\begin{figure}[htbp!]
    \centering
    \includegraphics[width=\hsize]{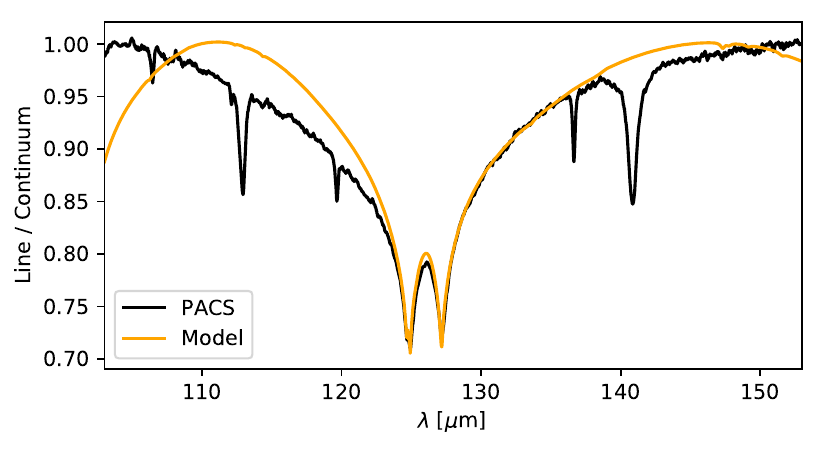}
    \caption{PACS data of the \ce{NH_3} $R(3)$ line along with a model spectrum obtained using $a_{\ce{NH_3}}(\infty)=1.5\times 10^{-5}$,  $p_\text{\ce{NH_3}}=0.7$\,bar, and $f_{\ce{NH_3}}=0.159$.}
    \label{fig:nh3_r3}
\end{figure}

We detected seven rotational bands of ammonia in the wavelength range of the PACS spectra: the \ce{NH_3} $R(2)$ to $R(8)$ bands. However, only the \ce{NH_3} $R(2)$, $R(3)$, $R(5)$, and $R(7)$ bands are unaffected by spectral leakage, so we only analyzed these. Out of these four bands, the \ce{NH_3} $R(3)$ band consistently showed large deviations from any model spectra. A separate analysis to find a best-fitting set of values for the ammonia abundance profile parameters for the \ce{NH_3} $R(3)$ band showed that its short-wavelength side ($\lambda<125\,\mu$m), especially the region between $100\,\mu$m and $110\,\mu$m did not agree with any model, while the long-wavelength side ($\lambda>125\,\mu$m) was fit well when suitable ammonia abundance profile parameters were applied (see Fig. \ref{fig:nh3_r3}). Therefore, we excluded the \ce{NH_3} $R(3)$ band from the ammonia analysis and all data between $100\,\mu$m and $110\,\mu$m from all further analyses. For a discussion on possible reasons for the misfit between the data and the models in the \ce{NH_3} $R(3)$ band, see Subsec. \ref{subsec:discussion_nh3}.

\begin{figure}[htbp!]
	\centering
	\includegraphics[width=\hsize]{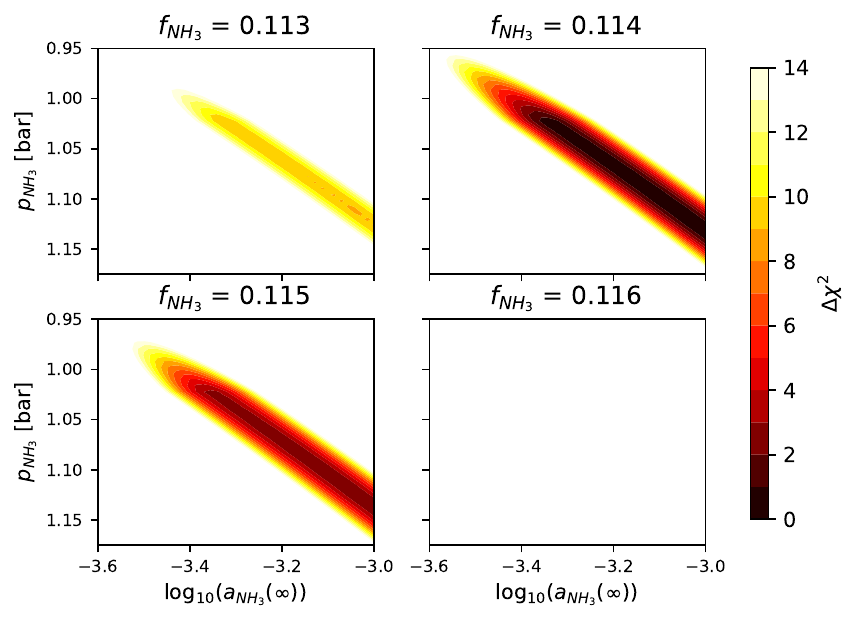}
    \caption{Least-squares comparison between the PACS data of the \ce{NH_3} $R(2)$, $R(5)$, and $R(7)$ bands and the model spectra synthesized using grids of ammonia abundance profiles expressed in $\Delta\chi^2$ (see Equation \ref{eq:delta_chi-squared}).}
	\label{fig:nh3_contours}
\end{figure}

Hence, we used the \ce{NH_3} $R(2)$, $R(5)$ and $R(7)$ bands and the a priori atmosphere model of Jupiter (see Sec. \ref{sec:model_setup}) with a family of ten thousand different ammonia abundance profiles employing $a_{\ce{NH_3}}(\infty)$ from $2\times 10^{-4}$ to $1\times 10^{-3}$, $p_{\ce{NH_3}}$ from 0.9\,bar to 1.2\,bar and $f_{\ce{NH_3}}$ from 0.113 to 0.116 to find the best-fitting parameter combination with the respective error bars. These limits on the parameters were set after an initial grid search with wider limits. Figure \ref{fig:nh3_contours} shows the least-squares results between the data and the model spectra using the \ce{NH_3} $R(2)$, $R(5)$, and $R(7)$ bands simultaneously. The least-squares results unveil a clear degeneracy between $a_{\ce{NH_3}}(\infty)$ and $p_{\ce{NH_3}}$, as a large range of combinations of these two parameters with $a_{\ce{NH_3}}(\infty)>5.3\times 10^{-4}$ and $p_{\ce{NH_3}}>1.04$\,bar deliver an equally good fit to the observations. Thus, we cannot constrain these two parameters beyond lower limits.

\begin{figure}[htbp!]
    \centering
    \includegraphics[width=\hsize]{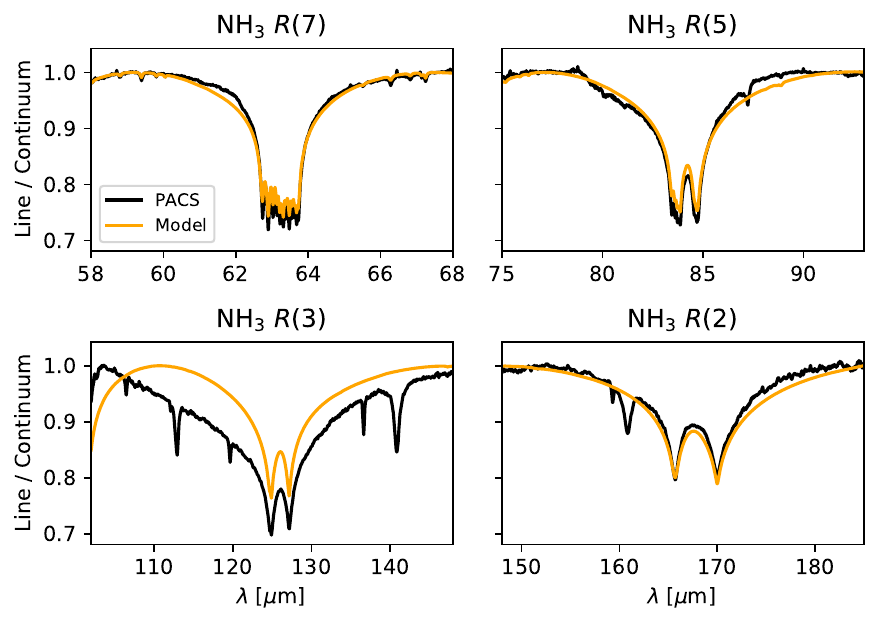}
    \caption{PACS data and best-fitting model spectra of the four ammonia bands that are considered in our analysis. The model spectra were calculated using $a_{\ce{NH_3}}(\infty)=5.3\times 10^{-4}$, $p_{\ce{NH_3}}=1.04$\,bar, and $f_{\ce{NH_3}}=0.114$.}
    \label{fig:nh3_lines}
\end{figure}

\begin{figure}[htbp!]
	\centering
	\includegraphics[width=\hsize]{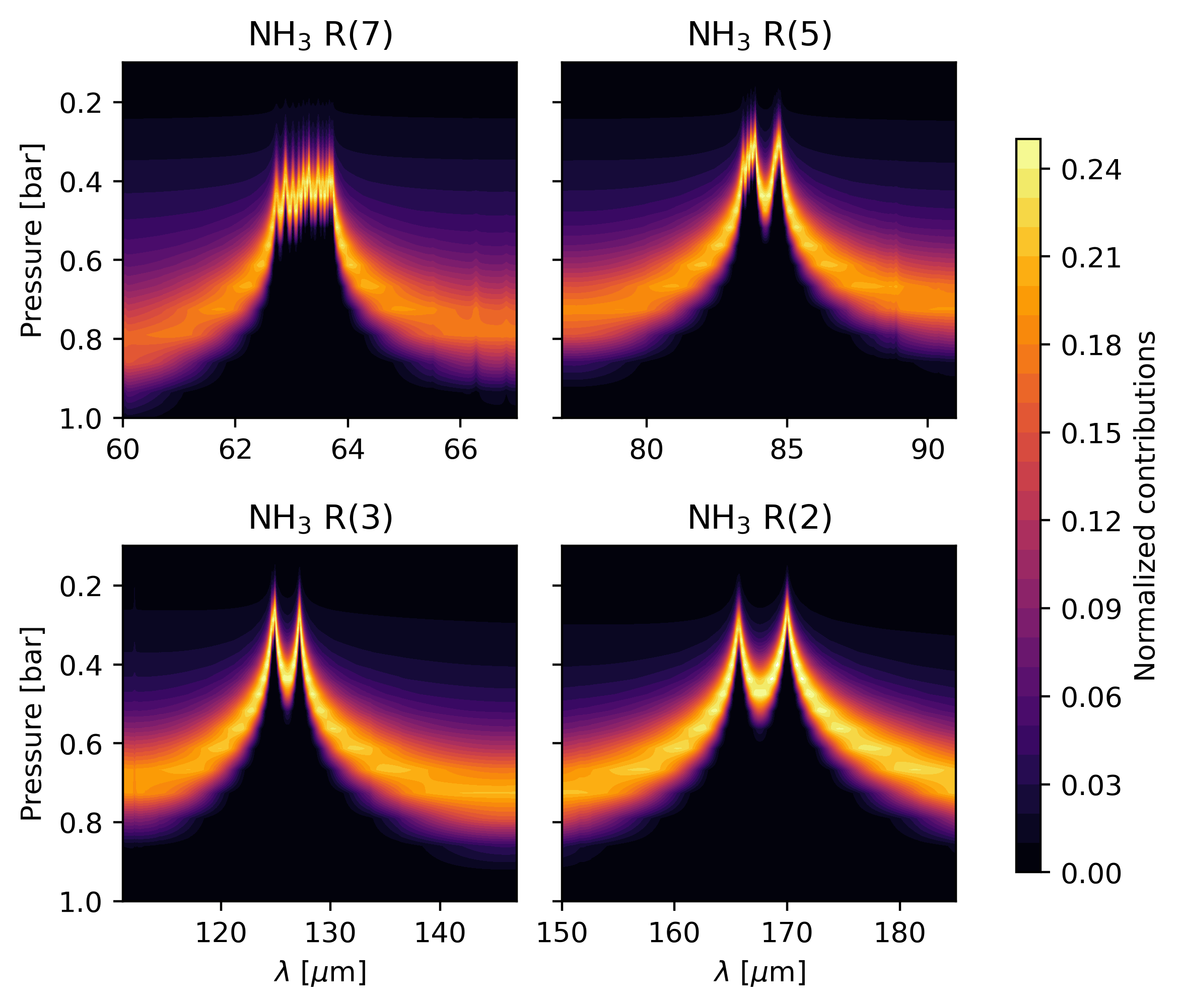}
	\caption{Normalized contribution functions for the observed ammonia bands derived from the same model as in Fig. \ref{fig:nh3_lines}.}
	\label{fig:nh3_contributions}
\end{figure}

The model spectra using $a_{\ce{NH_3}}(\infty)=5.3\times 10^{-4}$, $p_{\ce{NH_3}}=1.04$\,bar and $f_{\ce{NH_3}}=0.114$ as well as the PACS data of the four ammonia bands \ce{NH_3} $R(2)$, $R(3)$, $R(5)$, and $R(7)$ are shown in Fig. \ref{fig:nh3_lines}. We note that the \ce{NH_3} $R(3)$ band is plotted in that figure, even though we did not use it for the analysis of the ammonia profile in Jupiter.  The corresponding contribution functions of the forward models are presented in Fig. \ref{fig:nh3_contributions} and show that the observed absorption bands are sensitive to the ammonia mole fractions between $p\sim 275$\,mbar and $p\sim 900$\,mbar and thus to lower pressures than the knee pressure $p_{\ce{NH_3}}$ favored by the least-squares analysis (see Fig. \ref{fig:nh3_contours}). Thus, the degeneracy evident in Fig. \ref{fig:nh3_contours} originates in the fact that an increase of $a_{\ce{NH_3}}(\infty)$ with a simultaneous increase of $p_{\ce{NH_3}}$ can lead to the same ammonia abundances at pressures lower than $p_{\ce{NH_3}}$. Ammonia's fractional scale height is tightly constrained around $f_{\ce{NH_3}}=0.114$ and therefore, our analysis indicates a monotonic decrease of the ammonia abundance in the upper troposphere. The abundance profiles consistent with the observations within the $1\,\sigma$ range (see Subsec. \ref{subsec:discussion_nh3}) reveal that this decrease goes from an ammonia mole fraction of $(1.7\pm 0.8)\times 10^{-4}$ at $p\sim 900$\,mbar to $(1.7\pm 0.9)\times 10^{-8}$ at $p\sim 275$\,mbar.

\subsection{Phosphine} \label{subsec:ph3}
Phosphine does not participate in cloud condensation in Jupiter, but its abundance decreases in the upper troposphere due to photochemical reactions with solar photons. UV photons with wavelengths between 180\,nm and 235\,nm are required for the photodissociation of phosphine \citep{prinn75} and these photons penetrate the Jovian atmosphere up to about 100\,mbar \citep{irwin09}. Thus, similarly to the analysis of ammonia (Subsec. \ref{subsec:nh3}), we chose a three-parameter vertical abundance profile level of the shape
\begin{equation}
	a_{\ce{PH_3}}(p) =
	\begin{cases}
		a_{\ce{PH_3}}(\infty) & \text{if } p\ge p_{\ce{PH_3}} \\
		a_{\ce{PH_3}}(\infty)\left(\frac{p}{p_{\ce{PH_3}}}\right)^{((1-f_{\ce{PH_3}})/f_{\ce{PH_3}})} & \text{else} \label{eq:ph3_profile}
	\end{cases}
\end{equation}
for the analysis of the phosphine bands PACS observed. Here, $a_{\ce{PH_3}}(\infty)$ is the constant phosphine mole fraction below the knee pressure level $p_{\ce{PH_3}}$, above which the abundance depletes from the atmosphere. The parameter $f_{\ce{PH_3}}$ is phosphine's fractional scale height that governs the decrease of the phosphine abundance above the knee pressure level.

\begin{figure}[htbp!]
    \centering
    \includegraphics[width=\hsize]{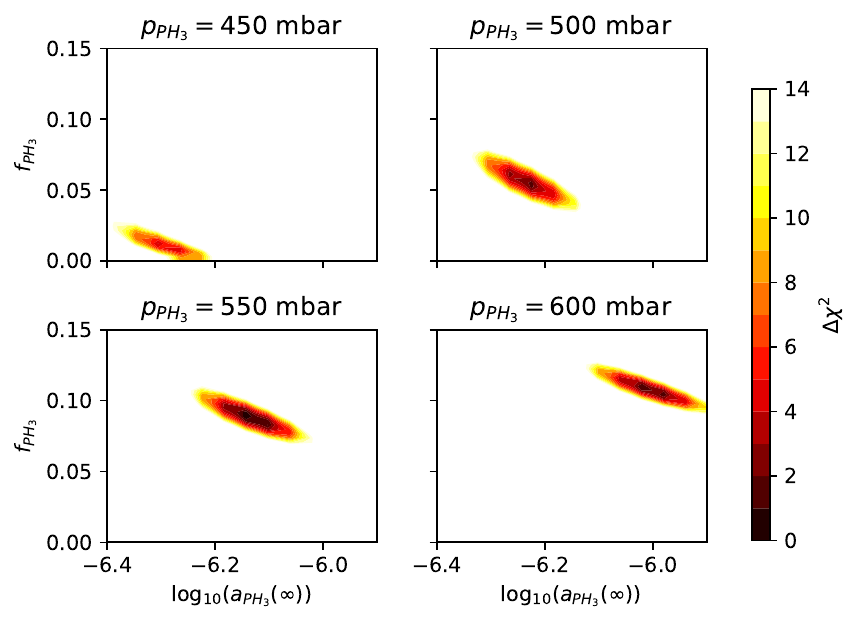}
    \caption{Least-squares results of the analysis of the \ce{PH_3} $R(6)$, $R(7)$, $R(9)$, $R(11)$, and $R(13)$ bands expressed using $\Delta\chi^2$ (Equation \ref{eq:delta_chi-squared}).}
    \label{fig:ph3_contours}
\end{figure}

Phosphine's rotational bands from \ce{PH_3} $R(5)$ to $R(22)$ lie in PACS's spectral range. However, the \ce{PH_3} $R(14)$ to $R(22)$ bands are undetectable in the PACS data due to their low amplitudes, the \ce{PH_3} $R(8)$ band falls into the center of the \ce{NH_3} $R(3)$ band and the \ce{PH_3} $R(5)$ and $R(10)$ bands suffer from PACS's spectral leakage. The \ce{PH_3} $R(5)$ band is only affected by spectral leakage at wavelengths longer than $\sim 188.3\,\mu$m, though. The \ce{PH_3} $R(12)$ band at $87.21\,\mu$m coincides with the \ce{CH_4} $R(10)$ band at $87.24\,\mu$m, creating one mutual spectral feature from $86.7\,\mu$m to $87.5\,\mu$m and we decided to use that feature for the analysis of the methane mole fraction in Jupiter (Subsec. \ref{subsec:ch4}) instead of the analysis of the phosphine abundance profile. Thus, we used the \ce{PH_3} $R(6)$, $R(7)$, $R(9)$, $R(11)$, and $R(13)$ bands for the analysis of phosphine in Jupiter's atmosphere. We ran the forward model simulations of the observed phosphine bands using the a priori model (see Sec. \ref{sec:model_setup}), the previously determined ammonia profile (see Subsec. \ref{subsec:nh3}) and a total of ten thousand phosphine abundance profiles employing $a_{\ce{PH_3}}(\infty)$ from $1\times 10^{-7}$ to $1\times 10^{-5}$, $p_{\ce{PH_3}}$ from $0.45$\,bar to $0.6$\,bar and $f_{\ce{PH_3}}$ from $1\times 10^{-4}$ to $0.15$. To reduce effects of the imperfect representation of the ammonia bands in the model spectra on the results of the phosphine analysis, we calculated line-to-continuum ratios of the phosphine bands with respect to the pseudo-continuum created by ammonia around the phosphine bands in both the model and the data. We approximated that pseudo-continuum by fitting second-order polynomials to the spectra outside of any spectral features except for the ammonia bands. The least-squares results for four different values for $p_{\ce{PH_3}}$ and grids of $f_{\ce{PH_3}}$ and $a_{\ce{PH_3}}(\infty)$ are plotted in Fig. \ref{fig:ph3_contours} using contours and point toward $a_{\ce{PH_3}}(\infty)=(7.2\pm1.2)\times 10^{-7}$, $p_{\ce{PH_3}}=(550\pm100)\,$mbar and $f_{\ce{PH_3}}=0.09\pm0.02$.

\begin{figure}[htbp!]
    \centering
    \includegraphics[width=\hsize]{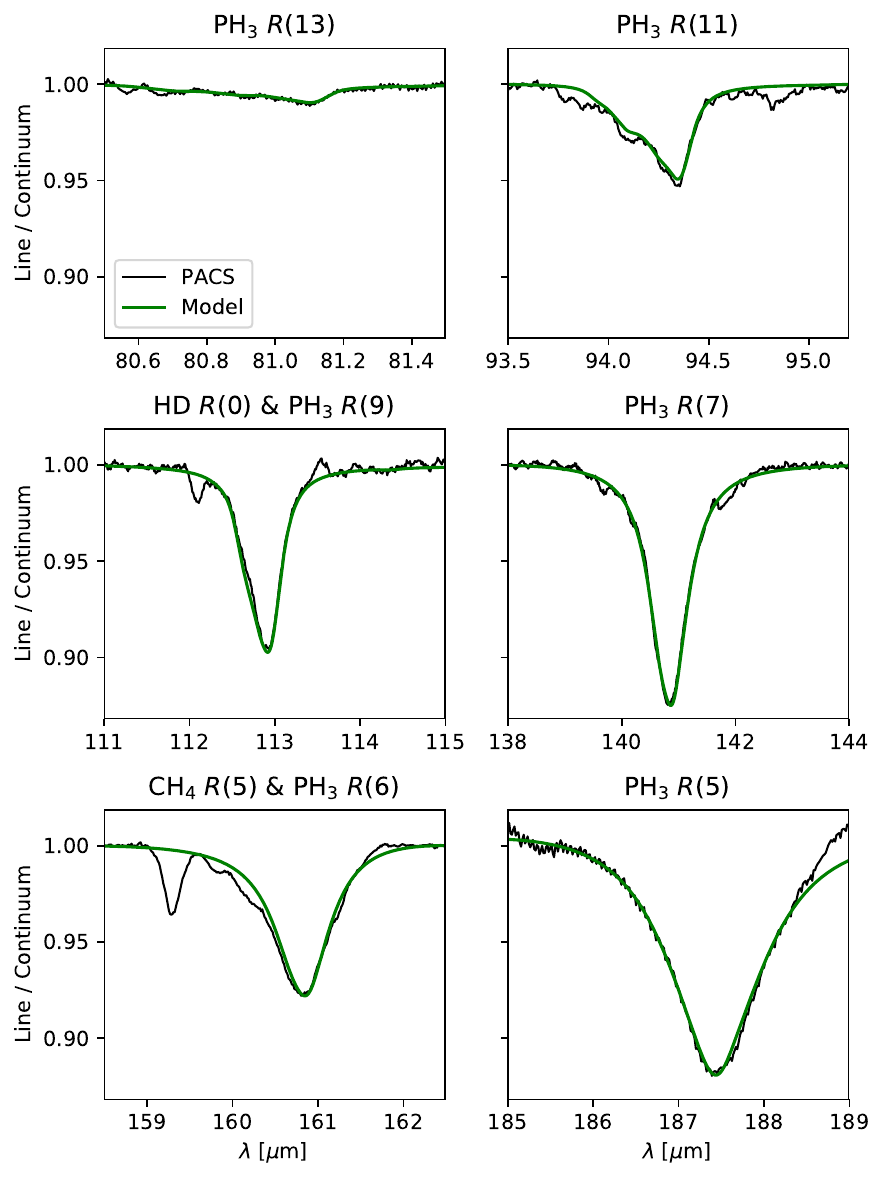}
    \caption{PACS data and model spectra of the five phosphine bands used for the analysis of phosphine in Jupiter's atmosphere as well as the \ce{PH_3} $R(5)$ band. The forward model spectra were synthesized using $a_{\ce{PH_3}}(\infty)=7.2\times 10^{-7}$, $p_{\ce{PH_3}}=550\,$mbar, and $f_{\ce{PH_3}}=0.09$.}
    \label{fig:ph3_lines}
\end{figure}

\begin{figure}[htbp!]
	\centering
	\includegraphics[width=\hsize]{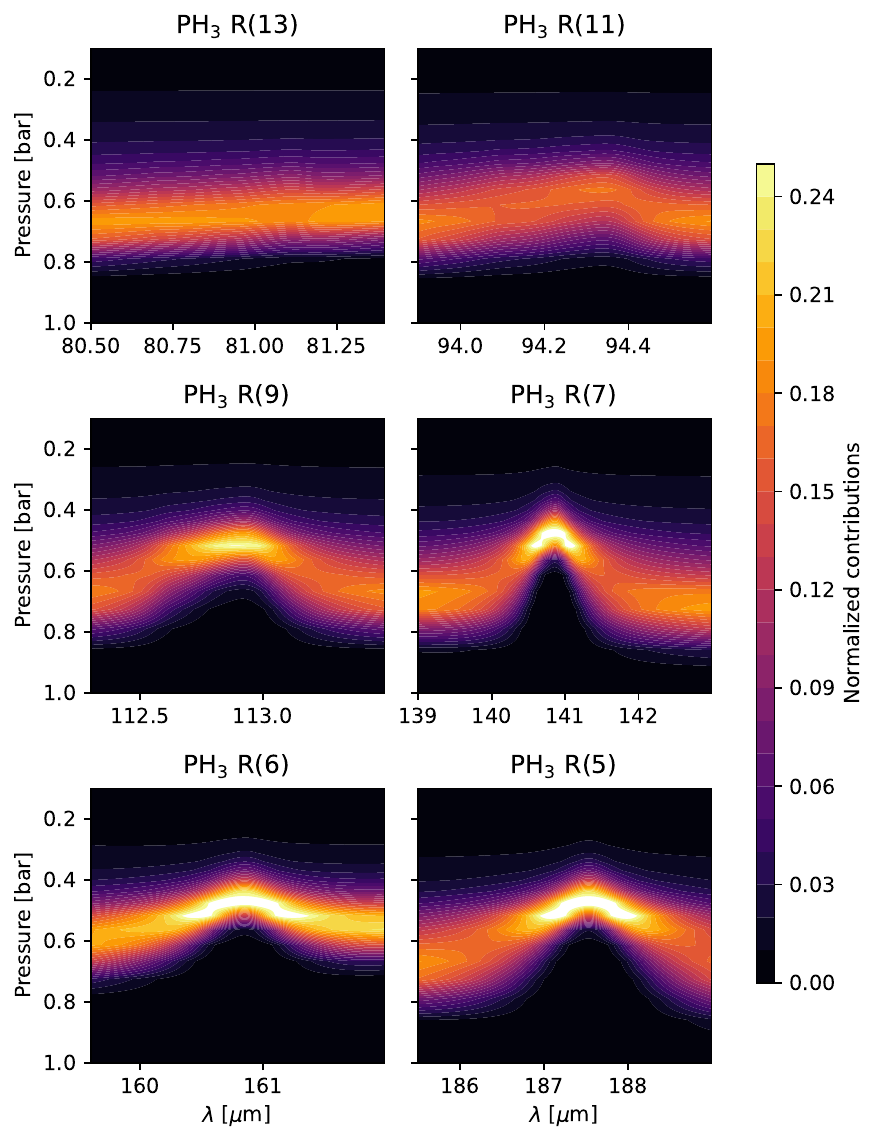}
	\caption{Normalized contribution functions for the observed phosphine bands derived from the same model as in Fig. \ref{fig:ph3_lines}.}
	\label{fig:ph3_contributions}
\end{figure}

The forward models of the absorption bands using the best-fitting parameters are plotted in Fig. \ref{fig:ph3_lines} along with the PACS observations. The models' corresponding contribution functions are shown in Fig. \ref{fig:ph3_contributions}. We note that the \ce{PH_3} $R(5)$ band is plotted in both these figures, even though we did not use it for the analysis of the phosphine profile in Jupiter. However, the phosphine profile found using the other five absorption bands performs well on the side of that band that is not affected by spectral leakage.

The contribution functions (Fig. \ref{fig:ph3_contributions}) show that the PACS observations are approximately sensitive to the phosphine abundances between $p\sim 350$\,mbar and $p\sim 850$\,mbar and our least-squares analysis returns $p_{\ce{PH_3}}=(550\pm100)\,$mbar for the knee pressure of the abundance profile which lies within the range of pressures probed by the observations. Therefore, our analysis of phosphine does not result in a degeneracy between $a_{\ce{PH_3}}(\infty)$ and $p_{\ce{PH_3}}$ as it was the case for $a_{\ce{NH_3}}(\infty)$ and $p_{\ce{NH_3}}$ in the ammonia analysis (see Subsec. \ref{subsec:nh3}). Thus, the data favor our chosen profile parameterization (Eq. \ref{eq:ph3_profile}) over a monotonic decrease of the phosphine abundance with height in the atmosphere and we are able to constrain all three parameters of our chosen abundance profile for phosphine.

\subsection{Methane} \label{subsec:ch4}
Methane does not condense into cloud particles in Jupiter and its photodissociation requires Solar UV photons that do not penetrate the Jovian atmosphere beyond $10^{-5}$\,bar \citep{moses05}. Thus, methane is stable in the part of Jupiter's atmosphere probed with PACS (see, e.g., Figs. \ref{fig:nh3_contributions} and \ref{fig:ph3_contributions}). Therefore, we assumed a vertically constant methane mole fraction for the analysis of methane in Jupiter's atmosphere.

PACS's spectral range includes the \ce{CH_4} $R(2)$ to $R(18)$ bands. However, the \ce{CH_4} $R(2)$ to $R(4)$ and the \ce{CH_4} $R(12)$ to $R(18)$ bands are undetectable in the PACS data due to their low amplitudes and the \ce{CH_4} $R(2)$ to $R(4)$ and $R(17)$ and $R(18)$ bands are additionally affected by spectral leakage. The \ce{CH_4} $R(16)$ band falls into the \ce{NH_3} $R(8)$ band center and the \ce{CH_4} $R(8)$ band at $106.5\,\mu$m lies within the troublesome region of the $R1$ spectral band that was identified in the analysis of the ammonia bands (Subsec. \ref{subsec:nh3}). As mentioned in the analysis of phosphine (Subsec. \ref{subsec:ph3}), the \ce{CH_4} $R(10)$ band coincides with the \ce{PH_3} $R(12)$ band which we did not use for the phosphine analysis. However, thanks to the good agreement between the PACS data and the model spectra for all other observed phosphine bands (see Fig. \ref{fig:ph3_lines}), we chose to use the spectral feature resulting from the \ce{PH_3} $R(12)$ and the \ce{CH_4} $R(10)$ bands for the analysis of Jupiter's atmospheric methane mole fraction. So, we used the \ce{CH_4} $R(5)$, $R(6)$, $R(7)$, $R(9)$, $R(10)$, and $R(11)$ bands simultaneously to infer the methane mole fraction in Jupiter's troposphere.

\begin{figure}[htbp!]
    \centering
    \includegraphics[width=\hsize]{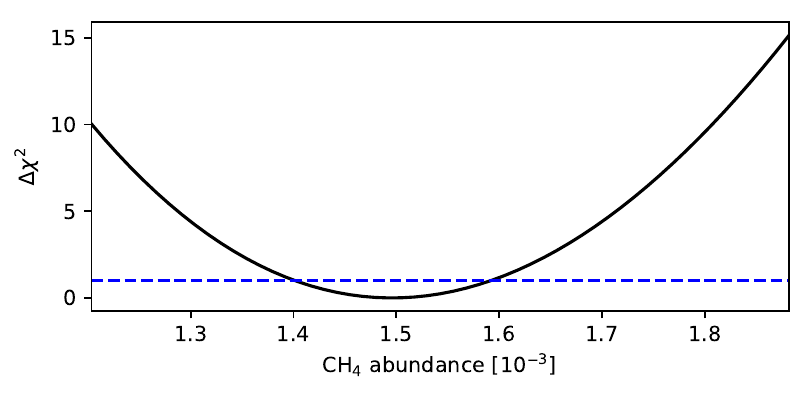}
    \caption{Results of the least-squares analysis of methane in Jupiter's atmosphere using the PACS data. The solid black line shows the $\Delta\chi^2$ values calculated for different methane mole fractions and the dashed blue line shows the $1\,\sigma$ confidence limit.}
    \label{fig:ch4_chisquared}
\end{figure}

\begin{figure}[htbp!]
    \centering
    \includegraphics[width=\hsize]{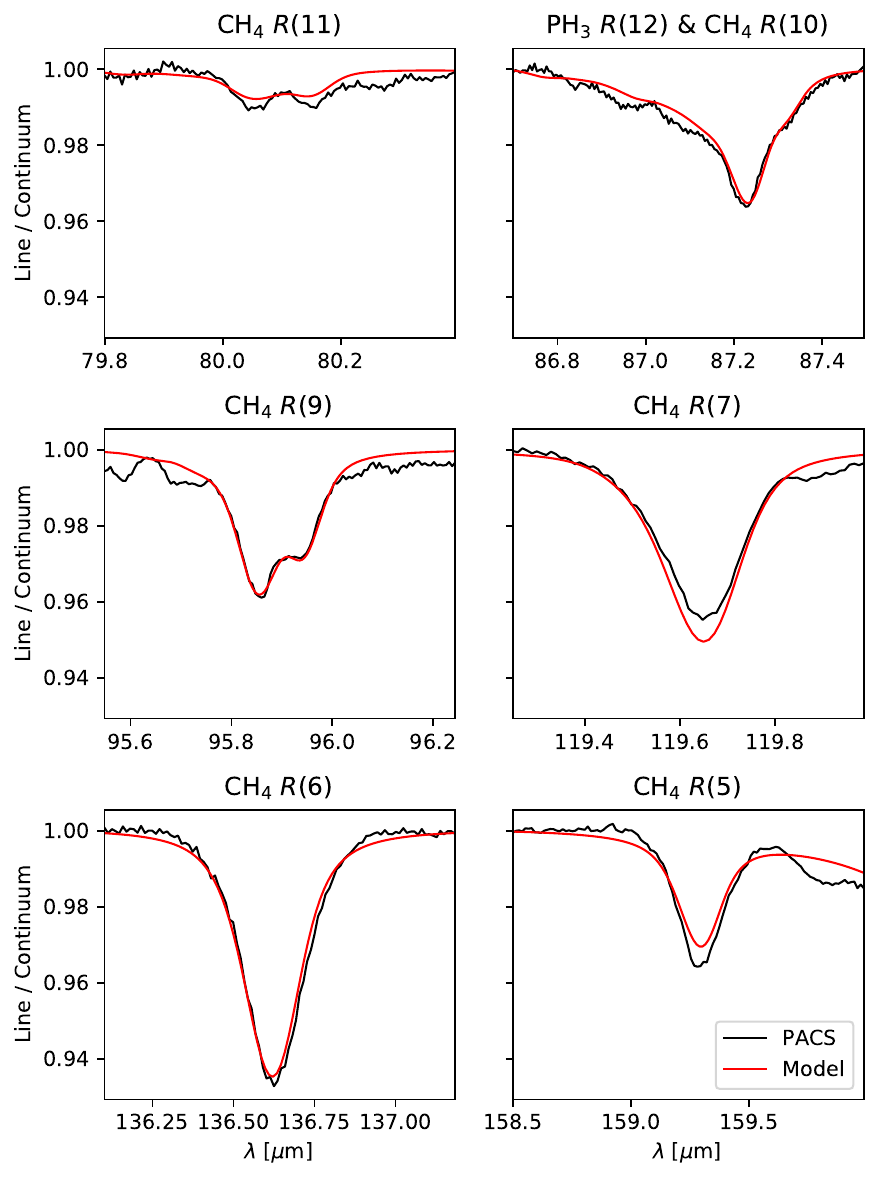}
    \caption{Methane bands we used to analyze the methane mole fraction in Jupiter's atmosphere along with their model spectra using a methane mole fraction of $1.49\times 10^{-3}$.}
    \label{fig:ch4_lines}
\end{figure}

\begin{figure}[htbp!]
	\centering
	\includegraphics[width=\hsize]{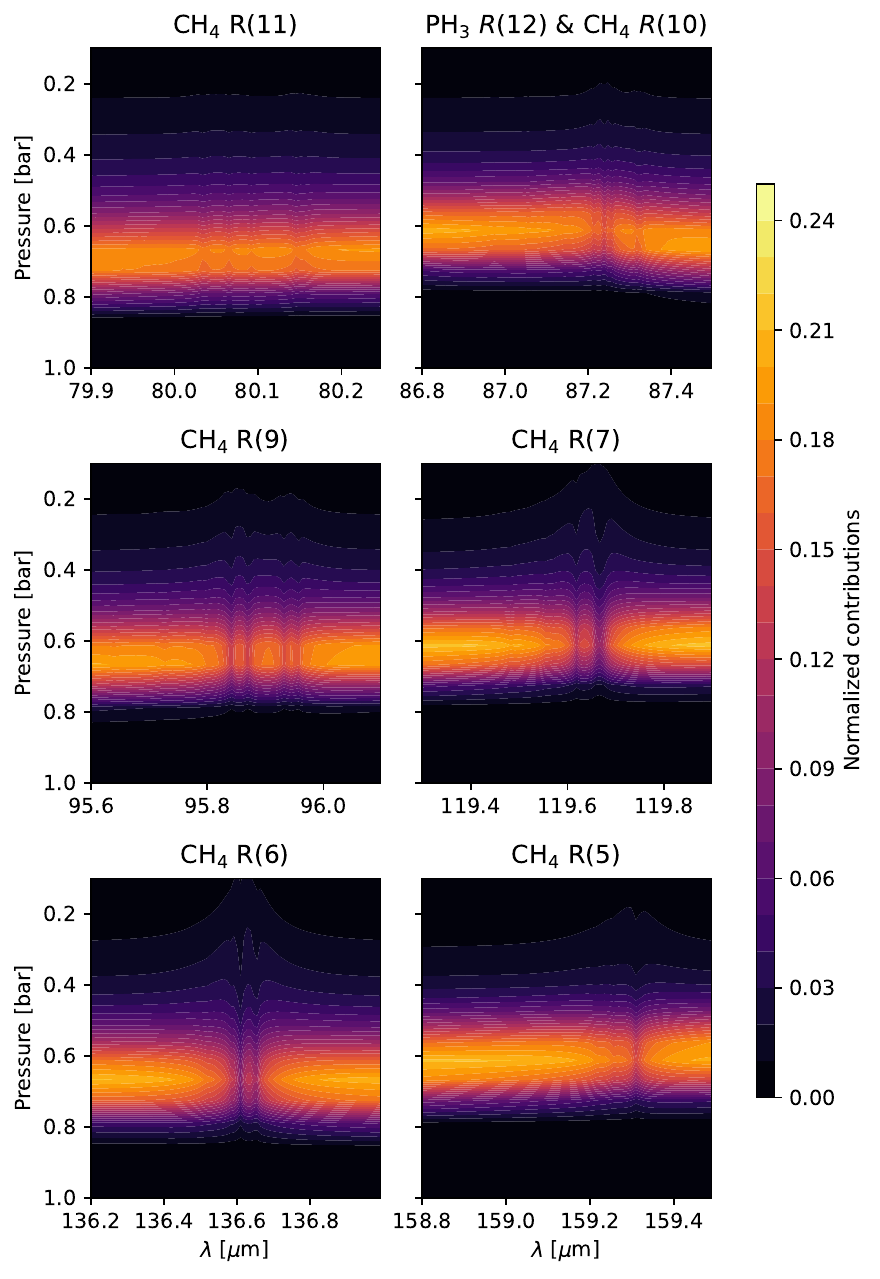}
	\caption{Normalized contribution functions for the observed methane bands derived from the same model as in Fig. \ref{fig:ch4_lines}.}
	\label{fig:ch4_contributions}
\end{figure}

The atmospheric model we used for the analysis of the methane bands consisted of the a priori atmosphere model (Sec. \ref{sec:model_setup}), the ammonia profile determined earlier (Subsec. \ref{subsec:nh3}), the previously found phosphine profile (Subsec. \ref{subsec:ph3}) and methane at different vertically homogeneous mole fractions. As in the analysis of phosphine (Subsec. \ref{subsec:ph3}), we calculated line-to-continuum ratios with respect to the pseudo-continuum created by ammonia in the Jovian atmosphere. We ran 200 forward model simulations of the above mentioned methane bands using methane mole fractions between $5\times 10^{-4}$ and $3\times 10^{-3}$ and inferred $(1.49\pm 0.09)\times 10^{-3}$ for the methane mole fraction in Jupiter's atmosphere using the least-squares method (see Fig. \ref{fig:ch4_chisquared}). The resulting best-fitting models and the PACS data of the methane bands are plotted in Fig. \ref{fig:ch4_lines}. The corresponding contribution functions of the models are plotted in Fig. \ref{fig:ch4_contributions} and show that the PACS observations are approximately sensitive to the pressure range from $p\sim 350$\,mbar to $p\sim 850$\,mbar.

\subsection{Deuterated hydrogen} \label{subsec:hd}
Deuterium is an outstandingly useful isotope in astrophysics and the Jovian \ce{D/H} ratio yields insight into the \ce{D/H} ratio of the protosolar nebula at the time of Jupiter's formation. Jupiter's atmospheric \ce{D/H} ratio can be inferred directly from the abundance of deuterated hydrogen in it.

\begin{figure}[htbp!]
    \centering
    \includegraphics[width=\hsize]{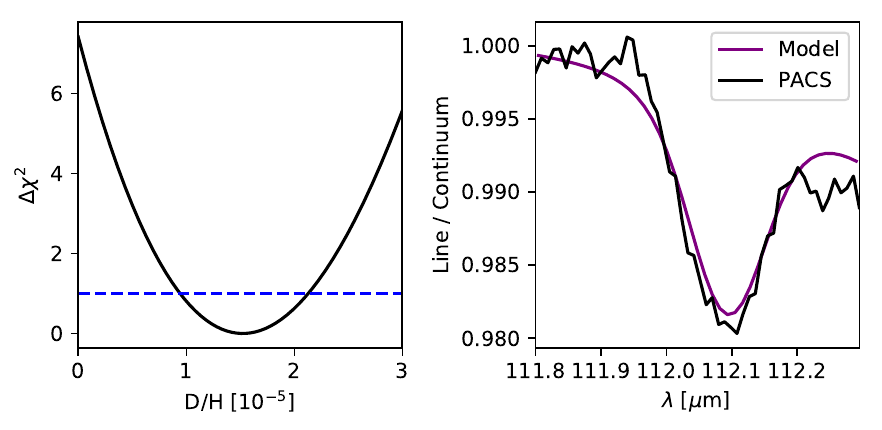}
    \caption{Results of the analysis of deuterated hydrogen. \textit{Left panel:} The least-squares results between the model spectra assuming a range of \ce{D/H} ratios and the PACS data expressed in $\Delta\chi^2$ (Equation \ref{eq:delta_chi-squared}) are plotted using a solid black line. The dashed blue line shows the $1\,\sigma$ confidence interval. \textit{Right panel:} The model spectrum using $\ce{D/H}=1.5\times 10^{-5}$ and the PACS data of the \ce{HD} $R(0)$ line.}
    \label{fig:hd_results}
\end{figure}

\begin{figure}[htbp!]
	\centering
	\includegraphics[width=\hsize]{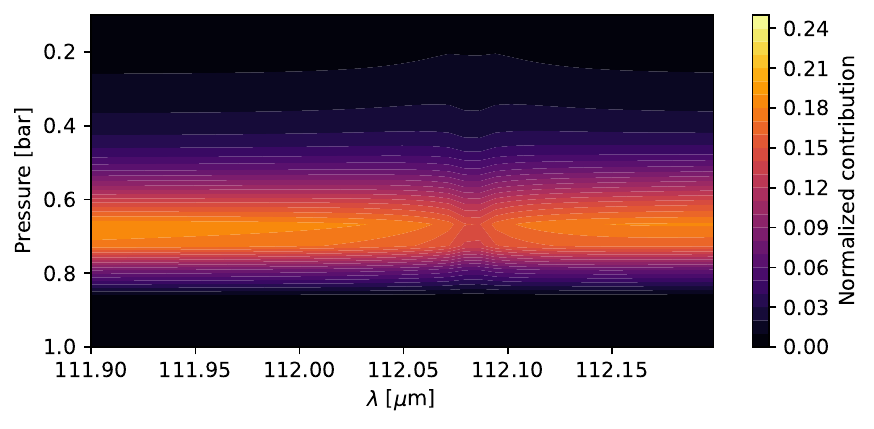}
	\caption{Normalized contribution functions for the observed HD $R(0)$ band using $\ce{D/H}=1.5\times 10^{-5}$.}
	\label{fig:hd_contributions}
\end{figure}

There are two rotational lines of deuterated hydrogen in the wavelength range of PACS: firstly, the \ce{HD} $R(1)$ line which is located at $56.2\,\mu$m, but this line is not visible in the PACS data, since it lies within the complex center of the \ce{NH_3} $R(8)$ band. Secondly, the \ce{HD} $R(0)$ line lies at $112.1\,\mu$m, nested into a wing of the \ce{PH_3} $R(9)$ band. Thus, we used the \ce{HD} $R(0)$ line to infer Jupiter's atmospheric \ce{D/H} ratio from the PACS data. Our forward model for the \ce{HD} analysis included the a priori Jupiter atmosphere (Sec. \ref{sec:model_setup}), the ammonia and phosphine profiles determined before (Subsecs. \ref{subsec:nh3} and \ref{subsec:ph3}), the methane mole fraction inferred earlier (Subsec. \ref{subsec:ch4}) and varying \ce{D/H} ratios. In total, we ran 1000 forward simulations using \ce{D/H} ratios between 0 and $3\times 10^{-5}$. As before, we normalized the PACS data and the model spectra to line-to-continuum ratios with respect to the ammonia spectrum approximated using a second-degree polynomial. The results of the least-squares analysis are shown in Fig. \ref{fig:hd_results} together with the model spectrum generated using the found best-fitting \ce{D/H} ratio and the PACS data of the \ce{HD} $R(0)$ line. The contribution functions corresponding to the best-fit forward model are plotted in Fig. \ref{fig:hd_contributions} and show that the observation is sensitive to the abundance of deuterated hydrogen between $p\sim 350$\,mbar and $p\sim 850$\,mbar in the atmosphere. Our analysis delivered a Jovian atmospheric \ce{D/H} ratio of $(1.5\pm0.6)\times 10^{-5}$.

\subsection{Hydrogen halides}
None of the hydrogen halides, hydrogen fluoride, hydrogen chloride, hydrogen bromide, and hydrogen iodide have been detected in Jupiter so far. \cite{showman01} suggested that their detection in the Jovian troposphere at temperatures lower than $400\,$K gets prevented by their reaction with ammonia to solid ammonium salts, which takes place on timescales several orders of magnitude shorter than the mean residence time of upper Jovian tropospheric air. In Jupiter's stratosphere, though, some of the hydrogen halides might be present due to influxes from external sources such as comets, the Galilean moon Io and Interplanetary Dust Particles (\citealp{fouchet04,teanby14}). However, a detection of any hydrogen halide in any part of Jupiter's atmosphere remains elusive.

Several rotational lines of the hydrogen halides lie in the spectral range of PACS. However, we detected none of them and thus, we inferred upper limits on their mole fractions instead. The a priori model of the Jovian atmosphere (see Sec. \ref{sec:model_setup}), the derived ammonia and phosphine profiles (Subsec. \ref{subsec:nh3} and \ref{subsec:ph3}) and the inferred methane mole fraction (Subsec. \ref{subsec:ch4}) were the base of the forward models for the following analyses. We assumed all four hydrogen halides to have a vertically constant abundance in Jupiter. All spectral lines they would have caused in the PACS spectra, if they had been present in observable quantities in the entire atmosphere, would have been absorption lines (see Figs. \ref{fig:hf}-\ref{fig:hi}). Thus, this analysis constrains the occurrence of hydrogen halides in the Jovian troposphere. To determine upper limits on their mole fractions, we varied them in the forward model one-by-one and compared each simulation with the PACS data around the strongest absorption lines of each molecule. The number of simulations and ranges of abundances for the forward model simulations varied from molecule to molecule and are given in the corresponding figures' captions. Our procedure to set upper limits (see Subsec. \ref{subsec:upper-limits}) delivered $3\sigma$ upper limits of $<1.1\times 10^{-11}$, $<6.0\times 10^{-11}$, $<2.3\times 10^{-10}$, and $<1.2\times 10^{-9}$ for the mole fractions of hydrogen fluoride (see Fig. \ref{fig:hf}), hydrogen chloride (see Fig. \ref{fig:hcl_tropo}), hydrogen bromide (see Fig. \ref{fig:hbr}), and hydrogen iodide (see Fig. \ref{fig:hi}), respectively.

Additionally, we derived an upper limit on the mole fraction of hydrogen chloride in the Jovian stratosphere by limiting its occurrence to pressures lower than $1\,$mbar in the forward model and varying its mole fraction that we assumed to be constant at pressures lower than that (similarly to \citealp{teanby14}). The same methodology as before delivered an upper limit of $<1.2\times 10^{-8}$ on the mole fraction of hydrogen chloride in Jupiter's stratosphere (see Fig. \ref{fig:hcl_strato}).

\section{Discussion} \label{sec:discussion}

\subsection{Ammonia} \label{subsec:discussion_nh3}
All ammonia bands observed with PACS are well represented by the same vertical ammonia distribution, except for the \ce{NH_3} $R(3)$ band (see Fig. \ref{fig:nh3_lines}). Neither the strength of that band nor the shape of its wing between $100$ and $110\,\mu$m agree between the model and the data, irrespective of the chosen ammonia abundance profile parameters (see Fig. \ref{fig:nh3_r3}). All observed ammonia bands probe very similar atmospheric pressure ranges (see Fig. \ref{fig:nh3_contributions}), and we therefore rule out any vertical variation of the abundance profile beyond our chosen parameterization (Eq. \ref{eq:nh3_profile}) as the cause of the mismatch. Another possible explanation for the mismatch is the horizontal heterogeneity of ammonia in Jupiter's troposphere driven by cloud condensation and the large-scale circulation of the atmosphere (see, e.g., \citealp{achterberg06,fletcher09,fletcher16,giles17,li17,dePater19,guillot20_2,guillot20_1,fletcher21,grassi21,moeckel23}). In the light of the analysis of the \ce{NH_3} $R(3)$ band which showed that the observations cannot be fit with any model using our profile parameterization (see Fig. \ref{fig:nh3_r3}), it is unlikely that the horizontal heterogeneity of ammonia in Jupiter's atmosphere caused the observed discrepancy. If the horizontal heterogeneity had caused the observed misfit of the \ce{NH_3} $R(3)$ band, another set of ammonia abundance profile parameters should be able to explain the observed spectrum. Thus, the disagreement between the model and the data at the short wavelength side of the \ce{NH_3} $R(3)$ band most likely stems from an instrumental or calibration issue of the PACS data. Therefore, we did not use the data between $100$ and $110\,\mu$m for the rest of the data analysis.

\begin{figure*}[htbp!]
    \centering
    \includegraphics[width=\hsize]{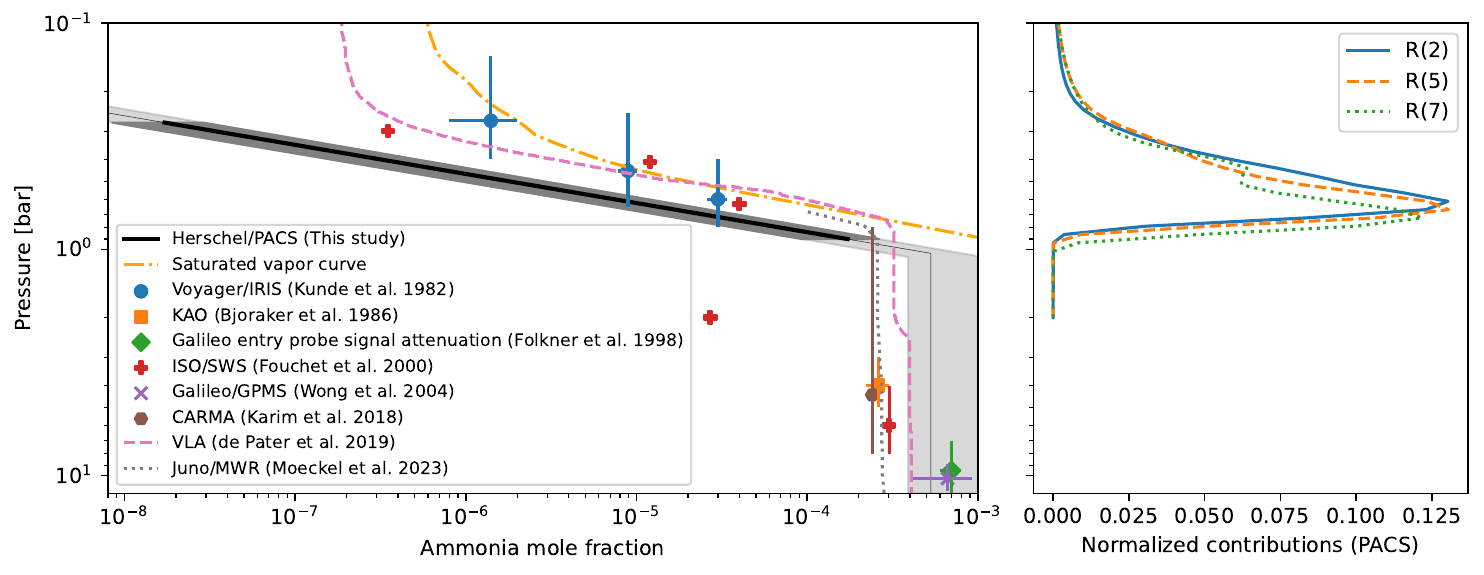}
    \caption{Ammonia mole fractions inferred from the PACS observations along with the results of previous studies, the saturated vapor curve, and the normalized contribution functions derived from our data analysis. \textit{Left panel:}  The thin solid line depicts the ammonia abundance profile derived from the PACS observations with light gray shading indicating the $1\,\sigma$ uncertainty range. The thick line segment with dark gray shading indicates the pressure range in which at least one of the normalized contribution functions (right panel) is $>0.02$ and thus presents the pressure range the PACS data are approximately sensitive to. The dash-dotted line represents the saturated vapor curve for ammonia calculated using Equation \ref{eq:nh3_saturation} and adopting our pressure-temperature profile. The colored data points represent results of previous studies with reported uncertainties and probed pressure levels indicated using horizontal and vertical errorbars, respectively. \textit{Right panel:} Normalized contribution functions integrated over the wavelength ranges of the five rotational bands used simultaneously to analyze the ammonia mole fractions in Jupiter using the PACS observations.}
    \label{fig:nh3}
\end{figure*}

Figure \ref{fig:nh3} presents the ammonia abundance profiles in the Jovian atmosphere that are consistent with our least-squares analysis within $1\,\sigma$ (Fig. \ref{fig:nh3_contours}), the contribution functions given by our model and the results of previous studies \citep{kunde82,bjoraker86,folkner98,fouchet00nh3,wong04,karim18,dePater19,moeckel23}. The contribution functions show that the observed absorption bands are sensitive to the ammonia mole fractions between $\sim 275$\,mbar and $\sim 900$\,mbar, which our least-squares analysis indicates to decrease from $(1.7\pm 0.8)\times 10^{-4}$ at $p\sim 900$\,mbar to $(1.7\pm 0.9)\times 10^{-8}$ at $p\sim 275$\,mbar, following a fractional scale height of $f_{\ce{NH_3}}\sim 0.114$ (see Subsec. \ref{subsec:nh3}). As the spectra we analyzed are averaged over the nine innermost spaxels (see Subsec. \ref{subsec:data_reduction}), these results correspond to a mean vertical distribution over a large part of the Jovian disk. Some of the previous studies whose results are indicated in Fig. \ref{fig:nh3} \citep{kunde82,folkner98,wong04} focused on individual locations on Jupiter, so discrepancies between our results and theirs might also stem from ammonia's latitudinal and longitudinal variability in the Jovian atmosphere.

To examine, whether or not our ammonia abundance profiles are consistent with ammonia cloud condensation in Jupiter, we calculated the saturated vapor curve for ammonia. The saturated vapor curve gives the mole fraction of ammonia at which the atmosphere were fully saturated and can be calculated using
\begin{equation}
    a_{\text{sat}}(p,T) = \frac{p_{\text{sub}}(T)}{p}, \label{eq:nh3_saturation}
\end{equation}
where $p$ and $T$ are the local atmospheric pressure and temperature and $p_{\text{sub}}(T)$ is the vapor pressure at sublimation which we calculated using
\begin{equation}
    p_{\text{sub}}(T) = \exp\left(15.96 - \frac{3537\,\text{K}}{T} + \mathcal{O}\left(T^{-2}\right)\right)\,\text{bar},
\end{equation}
where
\begin{equation}
    \mathcal{O}\left(T^{-2}\right) =  - \frac{3.31\cdot 10^{4}\,\text{K}^2}{T^2} + \frac{1.742\cdot 10^{6}\,\text{K}^3}{T^3} - \frac{2.995\cdot 10^{7}\,\text{K}^4}{T^4}
\end{equation}
\citep{fray09}. The saturated vapor curve is greater than the ammonia abundance profiles we derived from the PACS data in the entire atmosphere (see Fig. \ref{fig:nh3}). Thus, all ammonia mole fractions we inferred from the observations are subsaturated, suggesting that our disk-averaged spectra might be sensitive to dry, cloud-poor parts of the Jovian atmosphere. This might be caused by the fact that these regions emit more infrared radiation than cloudier parts of the atmosphere where the clouds absorb parts of Jupiter's thermal radiation.

\subsection{Phosphine}
The agreement between the PACS data and the modeled phosphine bands is excellent, since both the strengths and the shapes of all model spectra agree well with the PACS observations (see Fig. \ref{fig:ph3_lines}). Our least-squares analysis of the observed \ce{PH_3} $R(6)$, $R(7)$, $R(9)$, $R(11)$, and $R(13)$ bands returned values of $a_{\ce{PH_3}}(\infty)=(7.2\pm1.2)\times 10^{-7}$, $p_{\ce{PH_3}}=(550\pm100)\,$mbar, and $f_{\ce{PH_3}}=0.09\pm0.02$ for the chosen abundance profile's parameters (see Fig. \ref{fig:ph3_contours}) with an approximate sensitivity range from $p\sim 350$\,mbar to $p\sim 850$\,mbar (see Fig. \ref{fig:ph3_contributions}).

\begin{figure*}[htbp!]
    \centering
    \includegraphics[width=\hsize]{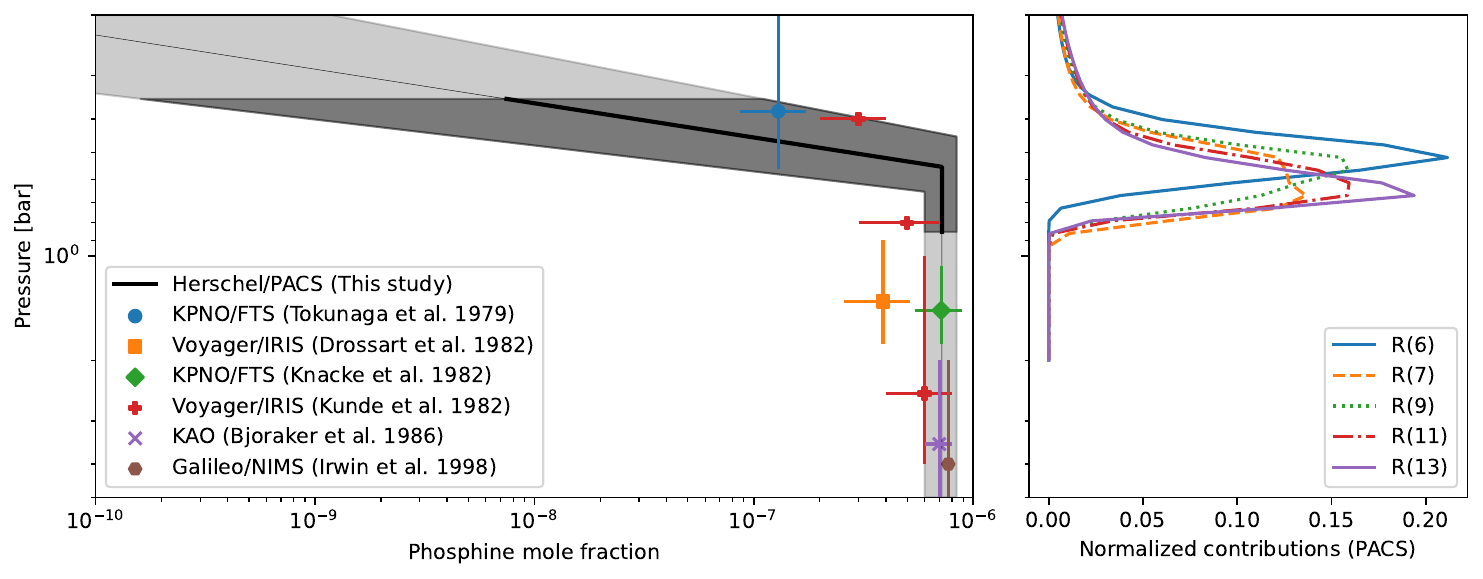}
    \caption{Same as Fig. \ref{fig:nh3}, but for phosphine and excluding a saturated vapor curve.}
    \label{fig:ph3}
\end{figure*}
Figure \ref{fig:ph3} shows the phosphine mole fractions inferred from our analysis along with the model's contribution functions integrated over the wavelength ranges of the five analyzed absorption bands and phosphine mole fractions determined in previous studies \citep{tokunaga79,drossart82,knacke82,kunde82,bjoraker86,irwin98}. The result for phosphine's mole fractions between $p\sim 550$\,mbar and $p\sim 850$\,mbar we inferred from the PACS data analysis agrees very well with the mole fractions in the lower troposphere determined using Voyager/IRIS \citep{kunde82}, the \textit{Kuiper} Airborne Observatory (KAO, \citealp{bjoraker86}) and \textit{Galileo}/NIMS \citep{irwin98}. This is evidence for very effective mixing of phosphine in the lower troposphere of Jupiter.

In addition to phosphine's mole fraction of $(7.7\pm 0.2)\times 10^{-7}$ in the $2-6$\,bar range, \cite{irwin98} inferred a fractional scale height of $f_{\ce{PH_3}}=0.271\pm0.014$ above 1\,bar which lies close to $f_{\ce{PH_3}}=0.3$ esimated by \cite{carlson93} from Voyager/IRIS data. Both our fractional scale height ($f_{\ce{PH_3}}=0.09\pm0.02$) and knee pressure ($p_{\ce{PH_3}}=(550\pm100)$\,mbar) are significantly lower than their respective values reported by \cite{irwin98}. However, our least-squares analysis demonstrates that with higher knee pressures, the corresponding best-fitting fractional scale height shifts to higher values (see Fig. \ref{fig:ph3_contours}). Since \cite{irwin98} prescribed phosphine's knee pressure in their analysis to 1\,bar, the same parameter correlation might have lead to an overestimation of the fractional scale height. Therefore, the discrepancy between our results and theirs might result from this assumption they made. Thus, the value for phosphine’s fractional scale height in Jupiter's upper troposphere might be lower than previously thought. The knee pressure we found lies closer to the 100\,mbar pressure level, up to which Solar UV photons that dissociate phosphine penetrate \citep{irwin09}. So, $p_{\ce{PH_3}}=(550\pm 100)$\,mbar might indeed be a more realistic value than the $1\,$bar assumed by \cite{irwin98}. Another possible reason for the differences between our fractional scale height result and theirs might be that phosphine is a molecule whose abundance profile significantly varies horizontally in Jupiter (see, e.g., \citealp{fletcher09,fletcher16}). \cite{irwin98} observed that the fractional scale height of phosphine decreased in dry, cloud-poor regions. The ammonia mole fractions inferred from the PACS observations suggest that the data might probe dry parts of the Jovian atmosphere (see Subsec. \ref{subsec:discussion_nh3}). Therefore, the fractional scale height of phosphine we inferred might be driven by our observations being especially sensitive to cloud-poor parts of Jupiter's atmosphere.

\subsection{Methane}
The methane bands observed with PACS are simultaneously well fit with the model using a methane mole fraction of $(1.49\pm 0.09)\times 10^{-3}$ (see Fig. \ref{fig:ch4_lines}). However, the strengths of the \ce{CH_4} $R(5)$, $R(7)$, and $R(11)$ bands in the PACS data deviate by up to 0.01 in line-to-continuum from the modeled strengths. As all absorption bands probe very similar pressure ranges in the atmosphere (see Fig. \ref{fig:ch4_contributions}), we exclude vertical variability of methane as a possible reason for the misfits. To examine, if the different band strengths might result from inaccurate line parameters of the HITRAN database, we ran the same analysis employing GEISA \citep{geisa} instead of HITRAN. However, this analysis returned the same methane mole fraction and the disagreement between the different line strengths remained. Therefore, if a line parameter problem exists, not only HITRAN is affected by it.

Table \ref{tab:ch4} lists the methane mole fraction inferred from the PACS data together with the results of previous investigations. The PACS result is consistent with the values found using Voyager/IRIS and the \textit{Galileo} entry probe, but it points toward a slightly lower mole fraction that corresponds to a Jovian \ce{C/H} ratio of $2.4\pm0.1$ times the protosolar value \citep{lodders19}. This ratio still represents a significant enrichment of carbon in Jupiter compared to the primordial composition and is in agreement with the core accretion theory \citep{pollack96}. The pressure ranges probed by the currently available observations do not indicate a significant variability of the methane mole fraction with height in Jupiter.
\begin{table*}[htbp!]
	\caption{Results for the methane mole fractions from previous studies and inferred from the PACS observations.}
	\label{tab:ch4}
	\centering
	\begin{tabular}{c c c c}
		\hline\hline
		\textbf{Data source} & \textbf{Mole fraction} & \textbf{Sensitivity range} & \textbf{References} \\ \hline
		Voyager/IRIS & $(1.69\pm0.19)\times 10^{-3}$ & \textit{'Upper troposphere'} & \cite{gautier82} \\
		KPNO/FTS & $(2.16\pm0.34)\times 10^{-3}$ & $p\sim(1.0-1.8)$\,bar & \cite{knacke82} \\
		KAO & $(3.0\pm1.0)\times 10^{-3}$ & $p\sim(1-6)$\,bar & \cite{bjoraker86} \\
		\textit{Galileo}/GPMS & $(1.81\pm0.34)\times 10^{-3}$ & $p\sim(0.5-3.8)$\,bar & \cite{niemann98} \\
		\textit{Galileo}/GPMS & $(2.04\pm 0.49)\times 10^{-3}$ & $p\sim(0.5-3.8)$\,bar & \cite{wong04} \\
		\textit{Herschel}/PACS & $(1.49\pm 0.09)\times 10^{-3}$ & $p\sim (0.35-0.8)$\,bar & This study \\ \hline
	\end{tabular}
\end{table*}

\subsection{D/H ratio}
The analysis of the \ce{HD} $R(0)$ line delivered $\ce{D/H}=(1.5\pm 0.6)\times 10^{-5}$ for Jupiter's atmosphere. Besides general uncertainties caused by calibration issues and imperfect modeling approaches, there are two additional, important sources of uncertainty for the size of the error bar of this result: Firstly, the value was derived using one spectral line only and secondly, that line is nested into a wing of the \ce{PH_3} $R(9)$ band. Fortunately, the agreement of the model spectra and the shape of the \ce{PH_3} bands is excellent (see Fig. \ref{fig:ph3_lines}). However, any uncertainties in the modeling of the \ce{PH_3} $R(9)$ band also affect the result of the \ce{D/H} ratio.

\begin{figure}[htbp!]
    \centering
	\includegraphics[width=\hsize]{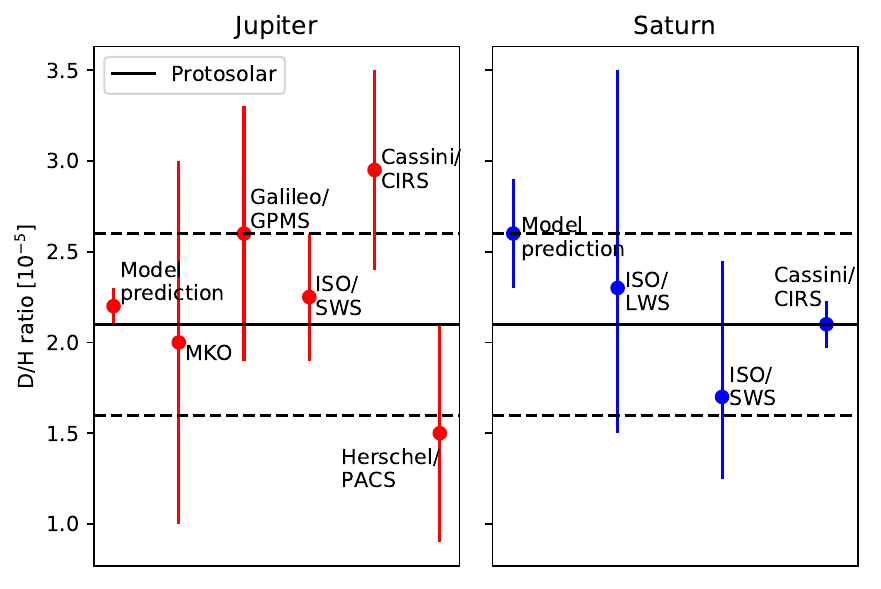}
	\caption{Values for the \ce{D/H} ratios of Jupiter and Saturn predicted by models of their interiors \citep{guillot99} and inferred from observations of \ce{HD} in their atmospheres. The \ce{D/H} ratio of the protosolar cloud inferred from the solar wind by \cite{geiss98} and its error bars are also indicated using a black solid line and black dashed lines, respectively. \textit{Left panel:} The values for the Jovian \ce{D/H} ratio based on observations with the Mauna Kea Observatory, \textit{Galileo}/GPMS, ISO/SWS and Cassini/CIRS were taken from \cite{smith89}, \cite{mahaffy98}, \cite{lellouch01} and \cite{pierel17}, respectively. \textit{Right panel:} The values for the Saturnian \ce{D/H} ratios inferred from ISO/LWS, ISO/SWS and Cassini/CIRS data were taken from \cite{griffin96}, \cite{lellouch01} and \cite{pierel17}, respectively.}
	\label{fig:d_h_compare}
\end{figure}

To interpret the meaning of our result, Fig. \ref{fig:d_h_compare} shows it along with all previous measurements of Jupiter's and Saturn's \ce{D/H} ratios from the abundances of \ce{HD} in their atmospheres, a measurement of the protosolar \ce{D/H} ratio derived from the solar wind and predictions derived from internal models of the two planets. Within the error bars, the Jovian \ce{D/H} ratio inferred from the PACS data is compatible with the protosolar ratio and all its previous determinations in Jupiter's atmosphere, except for the result of the Cassini/CIRS analysis. The internal models by \cite{guillot99} predicted a slightly greater atmospheric abundance of deuterium compared to the protosolar abundance in Saturn ($\ce{D/H}=2.6\pm 0.3$) than in Jupiter ($\ce{D/H}=2.2\pm 0.1$), that corresponds to a relative difference of less than $20\,\%$. However, the observationally determined \ce{D/H} ratios of the two planets are all afflicted with significant uncertainties on the order of about $10-50\,\%$ in $1\,\sigma$. Thus, it appears unlikely that the trend predicted by \cite{guillot99} is detectable based on the currently available data. We therefore argue that both planets are close representations of the protosolar \ce{D/H} ratio and that, up to today, no significant difference between their ratios can be concluded.

The PACS data demonstrate that the Jovian \ce{D/H} ratio could fall into the lower part of the range of possible primordial \ce{D/H} ratios and that it could be lower than Saturn's \ce{D/H} ratio (in contrast to the findings of \cite{lellouch01} and \cite{pierel17}, see Fig. \ref{fig:d_h_compare}). This suggests that icy planetesimals which were greatly enriched in deuterium (see, e.g., \citealp{lecluse96}) and probably played a vital role during Jupiter's formation did not accumulate deuterium in its atmosphere to a measurable degree. One possible explanation for this observation is that the planetesimals were not massive enough to modify the global \ce{D/H} ratio after mixing with the deuterium-poor protosolar hydrogen. Additionally, the exchange of mass between Jupiter's deep interior and envelope might not have been intense enough to increase the Jovian atmospheric \ce{D/H} ratio from the protosolar one. Indeed, \cite{guillot99} also found that Jupiter's atmospheric \ce{D/H} ratio should be enhanced by less than $10\,\%$ compared to the protosolar ratio. Thus, the PACS observation provides evidence in favor of \cite{guillot99}'s predictions.

\subsection{Hydrogen halides}
Several rotational lines of the hydrogen halides lie in PACS's spectral range and therefore, we were able to derive upper limits for hydrogen fluoride (\ce{HF}), hydrogen chloride (\ce{HCl}), hydrogen bromide (\ce{HBr}), and hydrogen iodide (\ce{HI}) in the Jovian troposphere. These upper limits are an improvement of up to two orders of magnitude compared to the formerly most stringent upper limits on their vertically constant assumed mole fractions determined using Cassini/CIRS data \citep{fouchet04}. Their upper limits, our new upper limits and the hydrogen halide mole fractions in Jupiter, if the ratios of the halogens' abundances to the hydrogen abundance in Jupiter were equal to those of the Sun, are listed in Table \ref{tab:hydrogen_halides_results}.
\begin{table*}[htbp!]
	\caption{$3\,\sigma$ upper limits for the mole fractions of hydrogen halides in the Jovian troposphere inferred from Cassini/CIRS observations \citep{fouchet04}, the PACS observations, and the molecules' respective Proto-Solar abundances.}
	\label{tab:hydrogen_halides_results}
	\centering
	\begin{tabular}{c c c c}
		\hline\hline
		\textbf{Molecule} & \makecell[c]{\textbf{New upper limit by PACS}} & \makecell[c]{\textbf{Previous upper limit} \\ \citep{fouchet04}} & \makecell[c]{\textbf{Proto-Solar composition mole fraction,} \\ \textbf{derived from} \cite{lodders19}} \\ \hline
		\ce{HF} & $<1.1\times 10^{-11}$ & $<2.7\times 10^{-11}$ & $(1.1\pm 0.2)\times 10^{-7}$ \\
		\ce{HCl} & $<6.0\times 10^{-11}$ & $<2.3\times 10^{-9}$ & $(3.6\pm 0.5)\times 10^{-7}$ \\
		\ce{HBr} & $<2.3\times 10^{-10}$ & $<1.0\times 10^{-9}$ & $(8.4\pm 1.7)\times 10^{-10}$ \\
		\ce{HI} & $<1.2\times 10^{-9}$ & $<7.6\times 10^{-9}$ & $(1.1\pm 0.4)\times  10^{-10}$ \\ \hline
	\end{tabular}
\end{table*}

\cite{fouchet04} ruled out a near-Solar or super-Solar abundance of hydrogen fluoride and hydrogen chloride in the Jovian troposphere, but they were not able to do so for hydrogen bromide and hydrogen iodide. For the first time, we can rule out a near- or super-Solar occurrence of hydrogen bromide in Jupiter's troposphere. Additionally, we reduced the previous upper limit on hydrogen iodide by a factor of about 7, but this value is still greater than its Solar abundance. This is further evidence that the hydrogen halides react with tropospheric ammonia to form solid ammonium salts and that this reaction is much faster than the vertical mixing time in Jupiter's troposphere \citep{showman01}.

The PACS data suggested an upper limit of $<1.2\times 10^{-8}$ on hydrogen chloride's mole fraction in the Jovian stratosphere. This limit is much higher than the one \cite{teanby14} inferred from \textit{Herschel}/HIFI data ($<6.1\times 10^{-11}$), so, we cannot improve this upper limit using the PACS data.

\section{Summary and conclusions}
The Jovian atmosphere was observed on October 31, 2009, using \textit{Herschel}/PACS which delivered high-quality FIR spectra between 50 and $220\,\mu$m. The PACS spectra contained absorption lines attributable to ammonia (\ce{NH_3}), methane (\ce{CH_4}), phosphine (\ce{PH_3}), and deuterated hydrogen (\ce{HD}) in the Jovian troposphere and emission lines attributable to water (\ce{H_2O}) in Jupiter's stratosphere. We used the water lines to calibrate the effective spectral resolution of PACS and then carried out an analysis of the disk-averaged spectra to infer the abundances of all other observed molecules in Jupiter's atmosphere. Additionally, we set upper limits on the mole fractions of the hydrogen halides.

Using the \ce{NH_3} $R(2)$, $R(5)$, and $R(7)$ bands, the mole fraction of ammonia was found to decrease from $(1.7\pm 0.8)\times 10^{-4}$ at $\sim 900$\,mbar to $(1.7\pm 0.9)\times 10^{-8}$ at $p\sim 275$\,mbar, following a fractional scale height of $f_{\ce{NH_3}}\sim 0.114$. For all other molecules, the PACS observations are sensitive to the pressure range between $p\sim 350$\,mbar and $p\sim 850$\,mbar. For phosphine, we found an atmospheric mole fraction of $(7.2\pm 1.2)\times 10^{-7}$ at pressures higher than $p_{\ce{PH_3}}=(550\pm 100)$\,mbar and a depletion following a fractional scale height of $f_{\ce{PH_3}}=(0.09\pm 0.02)$ at lower pressures. The inferred phosphine abundance in the lower troposphere agrees well with previous observations, but the fractional scale height is significantly lower than what has been determined previously. However, we found the resulting fractional scale height to strongly correlate with the knee pressure, underlining the importance of analyzing the impact of all three abundance profile parameters. Since previous studies held the knee pressure fixed at 1\,bar, our result suggests that the fractional scale height of phosphine in the upper troposphere might have been overestimated in the past. Our methane analysis delivered a mole fraction of $(1.49\pm 0.09)\times 10^{-3}$, which agrees well with previous results, but points toward a slightly lower, but still supersolar abundance of carbon in Jupiter ($\ce{C/H}=(2.4\pm 0.1)$ times the protosolar value). The model does not perfectly reproduce all of the six observed lines simultaneously using the best-fitting methane mole fraction and new measurements of methane lines might help to improve the fit. The \ce{HD} $R(0)$ line was detected in the PACS spectra and our analysis of that line delivered a new value for the Jovian \ce{D/H} ratio, $\ce{D/H}=(1.5\pm0.6)\times 10^{-5}$. This value agrees with previous observations of Jupiter's as well as the protosolar \ce{D/H} ratio inferred from the solar wind. It points toward a low \ce{D/H} ratio compared with the protosolar one, which suggests that
\begin{enumerate}
	\item the accretion of icy planetesimals into Jupiter during its formation did not significantly increase the Jovian atmospheric \ce{D/H} ratio, probably due to too little mass introduces by them as well as insufficient mass exchange between Jupiter's core and envelope \citep{guillot99}, and thus,
	\item the Jovian \ce{D/H} ratio is hence a good approximation for the protosolar ratio. 
\end{enumerate}
However, the size of the errorbars on the PACS measurement as well as all other currently available observations make a conclusive statement out of reach. Finally, the PACS data delivered new, more constraining upper limits for the mole fractions of hydrogen halides in Jupiter's troposphere. The new upper limits are $<1.1\times 10^{-11}$ for hydrogen fluoride, $<6.0\times 10^{-11}$ for hydrogen chloride , $<2.3\times 10^{-10}$ for hydrogen bromide, and $<1.2\times 10^{-9}$ for hydrogen iodide and are an improvement of up to two orders of magnitude compared to the previously most stringent upper limits on their tropospheric mole fractions inferred from Cassini/CIRS data \citep{fouchet04}. The new limits provided by PACS rule out near-Solar or super-Solar abundances of hydrogen fluoride, hydrogen chloride and, for the first time, hydrogen bromide in the Jovian troposphere. Thus, they provide further evidence for their condensation into solid ammonium salts in Jupiter's troposphere which was predicted to occur on much shorter timescales than vertical mixing  \citep{showman01}.

\section{Outlook}
The \textit{Herschel}/PACS observations carried out on October 31, 2009 helped to build a more complete picture of the Jovian spectrum at FIR wavelengths which advances the understanding of the chemical composition of Jupiter's atmosphere. One of their results is new insight into Jupiter's atmospheric \ce{D/H} ratio. This new finding can be further explored by future observations of Jupiter, as they are planned with, for example, the Submillimeter Wave Instrument (SWI) on board the Jupiter Icy Moons Explorer (JUICE, \citealp{juice_paper}). Using SWI, the abundance of \ce{HDO} in the Jovian stratosphere, and hence the \ce{D/H} ratio of Jupiter's stratospheric water, which primarily originates from the SL9 impact \citep{cavalie13}, will probably be measured. These prospective measurements might be able to shed light on the deuterium exchange mechanisms between water vapor and molecular hydrogen as well as the role of water in Jupiter's interior for its stratospheric water reservoir. This, in turn, might improve our understanding of the \ce{D/H} ratios of Jupiter, Saturn and the protosolar nebula, which are active topics of discussion to which the PACS observations contributed a new measurement.

\begin{acknowledgements}
The \textit{Herschel} spacecraft was designed, built, tested, and launched under a contract to ESA managed by the \textit{Herschel}/Planck Project team by an industrial consortium under the overall responsibility of the prime contractor Thales Alenia Space (Cannes), and including Astrium (Friedrichshafen) responsible for the payload module and for system testing at spacecraft level, Thales Alenia Space (Turin) responsible for the service module, and Astrium (Toulouse) responsible for the telescope, with in excess of a hundred subcontractors. PACS has been developed by a consortium of institutes led by MPE (Germany) and including UVIE (Austria); KU Leuven, CSL, IMEC (Belgium);  CEA, LAM (France); MPIA (Germany); INAF-IFSI/OAA/OAP/OAT, LENS, SISSA  (Italy); IAC (Spain). This development has been supported by the funding  agencies BMVIT (Austria), ESA-PRODEX (Belgium), CEA/CNES (France), DLR  (Germany), ASI/INAF (Italy), and CICYT/MCYT (Spain). We acknowledge the support of the DFG priority program SPP-1992 “Exploring the Diversity of ExtrasolarPlanets” (DFG PR 36 24602/41). This project would not have been possible without analysis tools which were made available online by the corresponding authors. The Planetary Spectrum Generator (PSG), JPL's HORIZONS web-interface, JPL Molecular Spectroscopy, HITRAN online and the GEISA database of molecular spectral lines were extraordinarily important for this study. We thank  the anonymous referee for the constructive feedback, which contributed to improving the quality of the manuscript.
\end{acknowledgements}

\bibliographystyle{aa}
\bibliography{literature}

\begin{thebibliography}{75}
\expandafter\ifx\csname natexlab\endcsname\relax\def\natexlab#1{#1}\fi

\bibitem[{Achterberg {et~al.}(2006)Achterberg, Conrath, \&
  Gierasch}]{achterberg06}
Achterberg, R.~K., Conrath, B.~J., \& Gierasch, P.~J. 2006, \icarus, 182, 169

\bibitem[{{Atreya} \& {Romani}(1985)}]{atreya85}
{Atreya}, S.~K. \& {Romani}, P.~N. 1985, in Recent Advances in Planetary
  Meteorology, ed. G.~E. {Hunt}, 17--68

\bibitem[{{Bjoraker} {et~al.}(1986){Bjoraker}, {Larson}, \&
  {Kunde}}]{bjoraker86}
{Bjoraker}, G.~L., {Larson}, H.~P., \& {Kunde}, V.~G. 1986, \icarus, 66, 579

\bibitem[{{Bjoraker} {et~al.}(2015){Bjoraker}, {Wong}, {de Pater}, \&
  {{\'A}d{\'a}mkovics}}]{bjoraker15}
{Bjoraker}, G.~L., {Wong}, M.~H., {de Pater}, I., \& {{\'A}d{\'a}mkovics}, M.
  2015, \apj, 810, 122

\bibitem[{{Burgdorf} {et~al.}(2001){Burgdorf}, {Encrenaz}, {Feuchtgruber},
  {Davis}, {Fouchet}, {Gautier}, {Lellouch}, {Orton}, \& {Sidher}}]{burgdorf02}
{Burgdorf}, M.~J., {Encrenaz}, T., {Feuchtgruber}, H., {et~al.} 2001, ESA
  Special Publication, 460, 365

\bibitem[{Carlson {et~al.}(1993)Carlson, Lacis, \& Rossow}]{carlson93}
Carlson, B.~E., Lacis, A.~A., \& Rossow, W.~B. 1993, \jgr, 98, 5251

\bibitem[{{Cavali{\'e}} {et~al.}(2021){Cavali{\'e}}, {Benmahi}, {Hue},
  {Moreno}, {Lellouch}, {Fouchet}, {Hartogh}, {Rezac}, {Greathouse},
  {Gladstone}, {Sinclair}, {Dobrijevic}, {Billebaud}, \& {Jarchow}}]{cavalie21}
{Cavali{\'e}}, T., {Benmahi}, B., {Hue}, V., {et~al.} 2021, \aap, 647, L8

\bibitem[{{Cavali{\'e}} {et~al.}(2008){Cavali{\'e}}, {Billebaud}, {Biver},
  {Dobrijevic}, {Lellouch}, {Brillet}, {Lecacheux}, {Hjalmarson}, {Sandqvist},
  {Frisk}, {Olberg}, {Bergin}, \& {Odin Team}}]{cavalie08}
{Cavali{\'e}}, T., {Billebaud}, F., {Biver}, N., {et~al.} 2008, \planss, 56,
  1573

\bibitem[{Cavali{\'e} {et~al.}(2013)Cavali{\'e}, Feuchtgruber, Lellouch,
  de~Val-Borro, Jarchow, Moreno, Hartogh, Orton, Greathouse, Billebaud,
  {et~al.}}]{cavalie13}
Cavali{\'e}, T., Feuchtgruber, H., Lellouch, E., {et~al.} 2013, \aap, 553, A21

\bibitem[{{de Pater} {et~al.}(2019{\natexlab{a}}){de Pater}, {Sault},
  {Moeckel}, {Moullet}, {Wong}, {Goullaud}, {DeBoer}, {Butler}, {Bjoraker},
  {{\'A}d{\'a}mkovics}, {Cosentino}, {Donnelly}, {Fletcher}, {Kasaba}, {Orton},
  {Rogers}, {Sinclair}, \& {Villard}}]{dePater19_ALMA}
{de Pater}, I., {Sault}, R.~J., {Moeckel}, C., {et~al.} 2019{\natexlab{a}},
  \aj, 158, 139

\bibitem[{{de Pater} {et~al.}(2019{\natexlab{b}}){de Pater}, {Sault}, {Wong},
  {Fletcher}, {DeBoer}, \& {Butler}}]{dePater19}
{de Pater}, I., {Sault}, R.~J., {Wong}, M.~H., {et~al.} 2019{\natexlab{b}},
  \icarus, 322, 168

\bibitem[{Drossart {et~al.}(1982)Drossart, Encrenaz, Kunde, Hanel, \&
  Combes}]{drossart82}
Drossart, P., Encrenaz, T., Kunde, V., Hanel, R., \& Combes, M. 1982, \icarus,
  49, 416

\bibitem[{{Encrenaz} {et~al.}(1996){Encrenaz}, {de Graauw}, {Schaeidt},
  {Lellouch}, {Feuchtgruber}, {Beintema}, {Bezard}, {Drossart}, {Griffin},
  {Heras}, {Kessler}, {Leech}, {Morris}, {Roelfsema}, {Roos-Serote}, {Salama},
  {Vandenbussche}, {Valentijn}, {Davis}, \& {Naylor}}]{encrenaz96}
{Encrenaz}, T., {de Graauw}, T., {Schaeidt}, S., {et~al.} 1996, \aap, 315, L397

\bibitem[{ESA(2013)}]{pacs_observers_manual}
ESA. 2013, {PACS Observers Manual}, 2nd edn.,
  \url{https://www.cosmos.esa.int/documents/12133/996891/PACS+Observers%27+Manual}

\bibitem[{Fletcher {et~al.}(2016)Fletcher, Greathouse, Orton, Sinclair, Giles,
  Irwin, \& Encrenaz}]{fletcher16}
Fletcher, L., Greathouse, T., Orton, G., {et~al.} 2016, \icarus, 278, 128

\bibitem[{Fletcher {et~al.}(2009)Fletcher, Orton, Yanamandra-Fisher, Fisher,
  Parrish, \& Irwin}]{fletcher09}
Fletcher, L., Orton, G., Yanamandra-Fisher, P., {et~al.} 2009, \icarus, 200,
  154

\bibitem[{{Fletcher} {et~al.}(2021){Fletcher}, {Oyafuso}, {Allison},
  {Ingersoll}, {Li}, {Kaspi}, {Galanti}, {Wong}, {Orton}, {Duer}, {Zhang},
  {Li}, {Guillot}, {Levin}, \& {Bolton}}]{fletcher21}
{Fletcher}, L.~N., {Oyafuso}, F.~A., {Allison}, M., {et~al.} 2021, \jgr, 126,
  e06858

\bibitem[{{Folkner} {et~al.}(1998){Folkner}, {Woo}, \& {Nandi}}]{folkner98}
{Folkner}, W.~M., {Woo}, R., \& {Nandi}, S. 1998, \jgr, 103, 22847

\bibitem[{Fouchet {et~al.}(2000{\natexlab{a}})Fouchet, Lellouch, B{\'e}zard,
  Encrenaz, Drossart, Feuchtgruber, \& de~Graauw}]{fouchet00nh3}
Fouchet, T., Lellouch, E., B{\'e}zard, B., {et~al.} 2000{\natexlab{a}},
  \icarus, 143, 223

\bibitem[{Fouchet {et~al.}(2000{\natexlab{b}})Fouchet, Lellouch, B{\'e}zard,
  Feuchtgruber, Drossart, \& Encrenaz}]{fouchet00hydro}
Fouchet, T., Lellouch, E., B{\'e}zard, B., {et~al.} 2000{\natexlab{b}}, \aap,
  355, L13

\bibitem[{Fouchet {et~al.}(2004)Fouchet, Orton, Irwin, Calcutt, \&
  Nixon}]{fouchet04}
Fouchet, T., Orton, G., Irwin, P.~G., Calcutt, S.~B., \& Nixon, C.~A. 2004,
  \icarus, 170, 237

\bibitem[{{Fray} \& {Schmitt}(2009)}]{fray09}
{Fray}, N. \& {Schmitt}, B. 2009, \planss, 57, 2053

\bibitem[{Gautier {et~al.}(1982)Gautier, Bezard, Marten, Baluteau, Scott,
  Chedin, Kunde, \& Hanel}]{gautier82}
Gautier, D., Bezard, B., Marten, A., {et~al.} 1982, \apj, 257, 901

\bibitem[{{Gautier} {et~al.}(1981){Gautier}, {Conrath}, {Flasar}, {Hanel},
  {Kunde}, {Chedin}, \& {Scott}}]{gautier81}
{Gautier}, D., {Conrath}, B., {Flasar}, M., {et~al.} 1981, \jgr, 86, 8713

\bibitem[{{Geiss} \& {Gloeckler}(1998)}]{geiss98}
{Geiss}, J. \& {Gloeckler}, G. 1998, \ssr, 84, 239

\bibitem[{{Giles} {et~al.}(2017){Giles}, {Fletcher}, {Irwin}, {Orton}, \&
  {Sinclair}}]{giles17}
{Giles}, R.~S., {Fletcher}, L.~N., {Irwin}, P. G.~J., {Orton}, G.~S., \&
  {Sinclair}, J.~A. 2017, \grl, 44, 10,838

\bibitem[{{Gordon} {et~al.}(2017){Gordon}, {Rothman}, {Hill}, {Kochanov},
  {Tan}, {Bernath}, {Birk}, {Boudon}, {Campargue}, {Chance}, {Drouin}, {Flaud},
  {Gamache}, {Hodges}, {Jacquemart}, {Perevalov}, {Perrin}, {Shine}, {Smith},
  {Tennyson}, {Toon}, {Tran}, {Tyuterev}, {Barbe}, {Cs{\'a}sz{\'a}r}, {Devi},
  {Furtenbacher}, {Harrison}, {Hartmann}, {Jolly}, {Johnson}, {Karman},
  {Kleiner}, {Kyuberis}, {Loos}, {Lyulin}, {Massie}, {Mikhailenko},
  {Moazzen-Ahmadi}, {M{\"u}ller}, {Naumenko}, {Nikitin}, {Polyansky}, {Rey},
  {Rotger}, {Sharpe}, {Sung}, {Starikova}, {Tashkun}, {Auwera}, {Wagner},
  {Wilzewski}, {Wcis{\l}o}, {Yu}, \& {Zak}}]{hitran}
{Gordon}, I.~E., {Rothman}, L.~S., {Hill}, C., {et~al.} 2017, \jqsrt, 203, 3

\bibitem[{Grasset {et~al.}(2013)Grasset, Dougherty, Coustenis, Bunce, Erd,
  Titov, Blanc, Coates, Drossart, Fletcher, {et~al.}}]{juice_paper}
Grasset, O., Dougherty, M., Coustenis, A., {et~al.} 2013, \planss, 78, 1

\bibitem[{Grassi {et~al.}(2017)Grassi, Adriani, Mura, Dinelli, Sindoni,
  Turrini, Filacchione, Migliorini, Moriconi, Tosi, {et~al.}}]{grassi17}
Grassi, D., Adriani, A., Mura, A., {et~al.} 2017, \grl, 44, 4615

\bibitem[{{Grassi} {et~al.}(2021){Grassi}, {Mura}, {Sindoni}, {Adriani},
  {Atreya}, {Filacchione}, {Fletcher}, {Lunine}, {Moriconi}, {Noschese},
  {Orton}, {Plainaki}, {Sordini}, {Tosi}, {Turrini}, {Olivieri},
  {Eichst{\"a}dt}, {Hansen}, {Melin}, {Altieri}, {Cicchetti}, {Dinelli},
  {Migliorini}, {Piccioni}, {Stefani}, \& {Bolton}}]{grassi21}
{Grassi}, D., {Mura}, A., {Sindoni}, G., {et~al.} 2021, \mnras, 503, 4892

\bibitem[{{Griffin} {et~al.}(1996){Griffin}, {Naylor}, {Davis}, {Ade},
  {Oldham}, {Swinyard}, {Gautier}, {Lellouch}, {Orton}, {Encrenaz}, {de
  Graauw}, {Furniss}, {Smith}, {Armand}, {Burgdorf}, {di Giorgio}, {Ewart},
  {Gry}, {King}, {Lim}, {Molinari}, {Price}, {Sidher}, {Smith}, {Texier},
  {Trams}, {Unger}, \& {Salama}}]{griffin96}
{Griffin}, M.~J., {Naylor}, D.~A., {Davis}, G.~R., {et~al.} 1996, \aap, 315,
  L389

\bibitem[{Guillot(1999)}]{guillot99}
Guillot, T. 1999, \planss, 47, 1183

\bibitem[{{Guillot} {et~al.}(2020{\natexlab{a}}){Guillot}, {Li}, {Bolton},
  {Brown}, {Ingersoll}, {Janssen}, {Levin}, {Lunine}, {Orton}, {Steffes}, \&
  {Stevenson}}]{guillot20_2}
{Guillot}, T., {Li}, C., {Bolton}, S.~J., {et~al.} 2020{\natexlab{a}}, \jgr,
  125, e06404

\bibitem[{{Guillot} {et~al.}(2020{\natexlab{b}}){Guillot}, {Stevenson},
  {Atreya}, {Bolton}, \& {Becker}}]{guillot20_1}
{Guillot}, T., {Stevenson}, D.~J., {Atreya}, S.~K., {Bolton}, S.~J., \&
  {Becker}, H.~N. 2020{\natexlab{b}}, \jgr, 125, e06403

\bibitem[{Hartogh {et~al.}(2009)Hartogh, Lellouch, Crovisier, Banaszkiewicz,
  Bensch, Bergin, Billebaud, Biver, Blake, Blecka, {et~al.}}]{hsso}
Hartogh, P., Lellouch, E., Crovisier, J., {et~al.} 2009, \planss, 57, 1596

\bibitem[{Heng \& Li(2021)}]{heng21}
Heng, K. \& Li, L. 2021, \apj Letters, 909, L20

\bibitem[{Irwin(2009)}]{irwin09}
Irwin, P. 2009, Giant planets of our solar system: atmospheres, composition,
  and structure (Springer Science \& Business Media)

\bibitem[{Irwin {et~al.}(1998)Irwin, Weir, Smith, Taylor, Lambert, Calcutt,
  Cameron-Smith, Carlson, Baines, Orton, {et~al.}}]{irwin98}
Irwin, P., Weir, A., Smith, S., {et~al.} 1998, \jgr, 103, 23001

\bibitem[{{Jacquinet-Husson} {et~al.}(2016){Jacquinet-Husson}, {Armante},
  {Scott}, {Ch{\'e}din}, {Cr{\'e}peau}, {Boutammine}, {Bouhdaoui},
  {Crevoisier}, {Capelle}, {Boonne}, {Poulet-Crovisier}, {Barbe}, {Chris
  Benner}, {Boudon}, {Brown}, {Buldyreva}, {Campargue}, {Coudert}, {Devi},
  {Down}, {Drouin}, {Fayt}, {Fittschen}, {Flaud}, {Gamache}, {Harrison},
  {Hill}, {Hodnebrog}, {Hu}, {Jacquemart}, {Jolly}, {Jim{\'e}nez},
  {Lavrentieva}, {Liu}, {Lodi}, {Lyulin}, {Massie}, {Mikhailenko},
  {M{\"u}ller}, {Naumenko}, {Nikitin}, {Nielsen}, {Orphal}, {Perevalov},
  {Perrin}, {Polovtseva}, {Predoi-Cross}, {Rotger}, {Ruth}, {Yu}, {Sung},
  {Tashkun}, {Tennyson}, {Tyuterev}, {Vander Auwera}, {Voronin}, \&
  {Makie}}]{geisa}
{Jacquinet-Husson}, N., {Armante}, R., {Scott}, N.~A., {et~al.} 2016, Journal
  of Molecular Spectroscopy, 327, 31

\bibitem[{{Karim} {et~al.}(2018){Karim}, {DeBoer}, {de Pater}, \&
  {Keating}}]{karim18}
{Karim}, R.~L., {DeBoer}, D., {de Pater}, I., \& {Keating}, G.~K. 2018, \aj,
  155, 129

\bibitem[{Knacke {et~al.}(1982)Knacke, Kim, Ridgway, \& Tokunaga}]{knacke82}
Knacke, R., Kim, S., Ridgway, S., \& Tokunaga, A. 1982, \apj, 262, 388

\bibitem[{{Kunde} {et~al.}(1982){Kunde}, {Hanel}, {Maguire}, {Gautier},
  {Baluteau}, {Marten}, {Chedin}, {Husson}, \& {Scott}}]{kunde82}
{Kunde}, V., {Hanel}, R., {Maguire}, W., {et~al.} 1982, \apj, 263, 443

\bibitem[{Lecluse {et~al.}(1996)Lecluse, Robert, Gautier, \&
  Guiraud}]{lecluse96}
Lecluse, C., Robert, F., Gautier, D., \& Guiraud, M. 1996, \planss, 44, 1579

\bibitem[{Lellouch {et~al.}(2001)Lellouch, B{\'e}zard, Fouchet, Feuchtgruber,
  Encrenaz, \& de~Graauw}]{lellouch01}
Lellouch, E., B{\'e}zard, B., Fouchet, T., {et~al.} 2001, \aap, 370, 610

\bibitem[{Lellouch {et~al.}(2002)Lellouch, B{\'e}zard, Moses, Davis, Drossart,
  Feuchtgruber, Bergin, Moreno, \& Encrenaz}]{lellouch02}
Lellouch, E., B{\'e}zard, B., Moses, J., {et~al.} 2002, \icarus, 159, 112

\bibitem[{Lewis(1969)}]{lewis69}
Lewis, J.~S. 1969, Icarus, 10, 365

\bibitem[{Li {et~al.}(2017)Li, Ingersoll, Janssen, Levin, Bolton, Adumitroaie,
  Allison, Arballo, Bellotti, Brown, {et~al.}}]{li17}
Li, C., Ingersoll, A., Janssen, M., {et~al.} 2017, \grl, 44, 5317

\bibitem[{{Lodders}(2019)}]{lodders19}
{Lodders}, K. 2019, arXiv e-prints, arXiv:1912.00844

\bibitem[{{Mahaffy} {et~al.}(1998){Mahaffy}, {Donahue}, {Atreya}, {Owen}, \&
  {Niemann}}]{mahaffy98}
{Mahaffy}, P.~R., {Donahue}, T.~M., {Atreya}, S.~K., {Owen}, T.~C., \&
  {Niemann}, H.~B. 1998, \ssr, 84, 251

\bibitem[{{Marten} {et~al.}(1995){Marten}, {Gautier}, {Griffin}, {Matthews},
  {Naylor}, {Davis}, {Owen}, {Orton}, {Bockel{\'e}e-Morvan}, {Colom},
  {Crovisier}, {Lellouch}, {de Pater}, {Atreya}, {Strobel}, {Han}, \&
  {Sanders}}]{marten95}
{Marten}, A., {Gautier}, D., {Griffin}, M.~J., {et~al.} 1995, \grl, 22, 1589

\bibitem[{{Moeckel} {et~al.}(2023){Moeckel}, {de Pater}, \&
  {DeBoer}}]{moeckel23}
{Moeckel}, C., {de Pater}, I., \& {DeBoer}, D. 2023, The Planetary Science
  Journal, 4, 25

\bibitem[{{Moreno} {et~al.}(2003){Moreno}, {Marten}, {Matthews}, \&
  {Biraud}}]{moreno03}
{Moreno}, R., {Marten}, A., {Matthews}, H.~E., \& {Biraud}, Y. 2003, \planss,
  51, 591

\bibitem[{Moses {et~al.}(2005)Moses, Fouchet, B{\'e}zard, Gladstone, Lellouch,
  \& Feuchtgruber}]{moses05}
Moses, J., Fouchet, T., B{\'e}zard, B., {et~al.} 2005, \jgr, 110

\bibitem[{{Niemann} {et~al.}(1998){Niemann}, {Atreya}, {Carignan}, {Donahue},
  {Haberman}, {Harpold}, {Hartle}, {Hunten}, {Kasprzak}, {Mahaffy}, {Owen}, \&
  {Way}}]{niemann98}
{Niemann}, H.~B., {Atreya}, S.~K., {Carignan}, G.~R., {et~al.} 1998, \jgr, 103,
  22831

\bibitem[{Nixon {et~al.}(2007)Nixon, Achterberg, Conrath, Irwin, Teanby,
  Fouchet, Parrish, Romani, Abbas, LeClair, {et~al.}}]{nixon07}
Nixon, C., Achterberg, R., Conrath, B., {et~al.} 2007, \icarus, 188, 47

\bibitem[{Nixon {et~al.}(2010)Nixon, Achterberg, Romani, Allen, Zhang, Teanby,
  Irwin, \& Flasar}]{nixon10}
Nixon, C.~A., Achterberg, R.~K., Romani, P.~N., {et~al.} 2010, \planss, 58,
  1667

\bibitem[{Noll(1996)}]{noll96}
Noll, K.~S. 1996, \icarus, 124, 608

\bibitem[{{Pierel} {et~al.}(2017){Pierel}, {Nixon}, {Lellouch}, {Fletcher},
  {Bjoraker}, {Achterberg}, {B{\'e}zard}, {Hesman}, {Irwin}, \&
  {Flasar}}]{pierel17}
{Pierel}, J.~D.~R., {Nixon}, C.~A., {Lellouch}, E., {et~al.} 2017, \aj, 154,
  178

\bibitem[{Pilbratt {et~al.}(2010)Pilbratt, Riedinger, Passvogel, Crone, Doyle,
  Gageur, Heras, Jewell, Metcalfe, Ott, {et~al.}}]{herschel1}
Pilbratt, G., Riedinger, J., Passvogel, T., {et~al.} 2010, \aap, 518, L1

\bibitem[{{Poglitsch} {et~al.}(2008){Poglitsch}, {Waelkens}, {Bauer}, {Cepa},
  {Feuchtgruber}, {Henning}, {van Hoof}, {Kerschbaum}, {Krause}, {Renotte},
  {Rodriguez}, {Saraceno}, \& {Vandenbussche}}]{pacs08}
{Poglitsch}, A., {Waelkens}, C., {Bauer}, O.~H., {et~al.} 2008, in Society of
  Photo-Optical Instrumentation Engineers (SPIE) Conference Series, Vol. 7010,
  Space Telescopes and Instrumentation 2008: Optical, Infrared, and Millimeter,
  ed. J.~{Oschmann}, Jacobus~M., M.~W.~M. {de Graauw}, \& H.~A. {MacEwen},
  701005

\bibitem[{Poglitsch {et~al.}(2010)Poglitsch, Waelkens, Geis, Feuchtgruber,
  Vandenbussche, Rodriguez, Krause, Renotte, Van~Hoof, Saraceno,
  {et~al.}}]{pacs10}
Poglitsch, A., Waelkens, C., Geis, N., {et~al.} 2010, \aap, 518, L2

\bibitem[{Pollack {et~al.}(1996)Pollack, Hubickyj, Bodenheimer, Lissauer,
  Podolak, \& Greenzweig}]{pollack96}
Pollack, J.~B., Hubickyj, O., Bodenheimer, P., {et~al.} 1996, \icarus, 124, 62

\bibitem[{Press {et~al.}(2007)Press, Teukolsky, Vetterling, \& Flannery}]{nr}
Press, W.~H., Teukolsky, S.~A., Vetterling, W.~T., \& Flannery, B.~P. 2007,
  {Numerical Recipes: The Art of Scientific Computing}, 3rd edn. (Cambridge
  University Press)

\bibitem[{{Prinn} \& {Lewis}(1975)}]{prinn75}
{Prinn}, R.~G. \& {Lewis}, J.~S. 1975, Science, 190, 274

\bibitem[{{Ragent} {et~al.}(1998){Ragent}, {Colburn}, {Rages}, {Knight},
  {Avrin}, {Orton}, {Yanamandra-Fisher}, \& {Grams}}]{galileo99clouds}
{Ragent}, B., {Colburn}, D.~S., {Rages}, K.~A., {et~al.} 1998, \jgr, 103, 22891

\bibitem[{Roos-Serote {et~al.}(2004)Roos-Serote, Atreya, Wong, \&
  Drossart}]{roos04}
Roos-Serote, M., Atreya, S., Wong, M., \& Drossart, P. 2004, \planss, 52, 397

\bibitem[{{Seiff} {et~al.}(1998){Seiff}, {Kirk}, {Knight}, {Young}, {Mihalov},
  {Young}, {Milos}, {Schubert}, {Blanchard}, \& {Atkinson}}]{seiff98_galileo}
{Seiff}, A., {Kirk}, D.~B., {Knight}, T. C.~D., {et~al.} 1998, \jgr, 103, 22857

\bibitem[{Showman(2001)}]{showman01}
Showman, A.~P. 2001, \icarus, 152, 140

\bibitem[{Smith {et~al.}(1989)Smith, Schempp, \& Baines}]{smith89}
Smith, W.~H., Schempp, W., \& Baines, K.~H. 1989, \apj, 336, 967

\bibitem[{Teanby {et~al.}(2014)Teanby, Showman, Fletcher, \& Irwin}]{teanby14}
Teanby, N., Showman, A., Fletcher, L., \& Irwin, P. 2014, \planss, 103, 250

\bibitem[{{Tokunaga} {et~al.}(1979){Tokunaga}, {Ridgway}, {Wallace}, \&
  {Knacke}}]{tokunaga79}
{Tokunaga}, A.~T., {Ridgway}, S.~T., {Wallace}, L., \& {Knacke}, R.~F. 1979,
  \apj, 232, 603

\bibitem[{{Villanueva} {et~al.}(2018){Villanueva}, {Smith}, {Protopapa},
  {Faggi}, \& {Mandell}}]{psg}
{Villanueva}, G.~L., {Smith}, M.~D., {Protopapa}, S., {Faggi}, S., \&
  {Mandell}, A.~M. 2018, \jqsrt, 217, 86

\bibitem[{Von~Zahn \& Hunten(1996)}]{zahn96}
Von~Zahn, U. \& Hunten, D. 1996, Science, 272, 849

\bibitem[{Weidenschilling \& Lewis(1973)}]{weidenschilling73}
Weidenschilling, S. \& Lewis, J. 1973, Icarus, 20, 465

\bibitem[{Wong {et~al.}(2004)Wong, Mahaffy, Atreya, Niemann, \& Owen}]{wong04}
Wong, M.~H., Mahaffy, P.~R., Atreya, S.~K., Niemann, H.~B., \& Owen, T.~C.
  2004, \icarus, 171, 153

\end{thebibliography}

\begin{appendix}
\section{Hydrogen halide upper limits}
\begin{figure}[!htbp]
	\centering
	\includegraphics[width=\hsize]{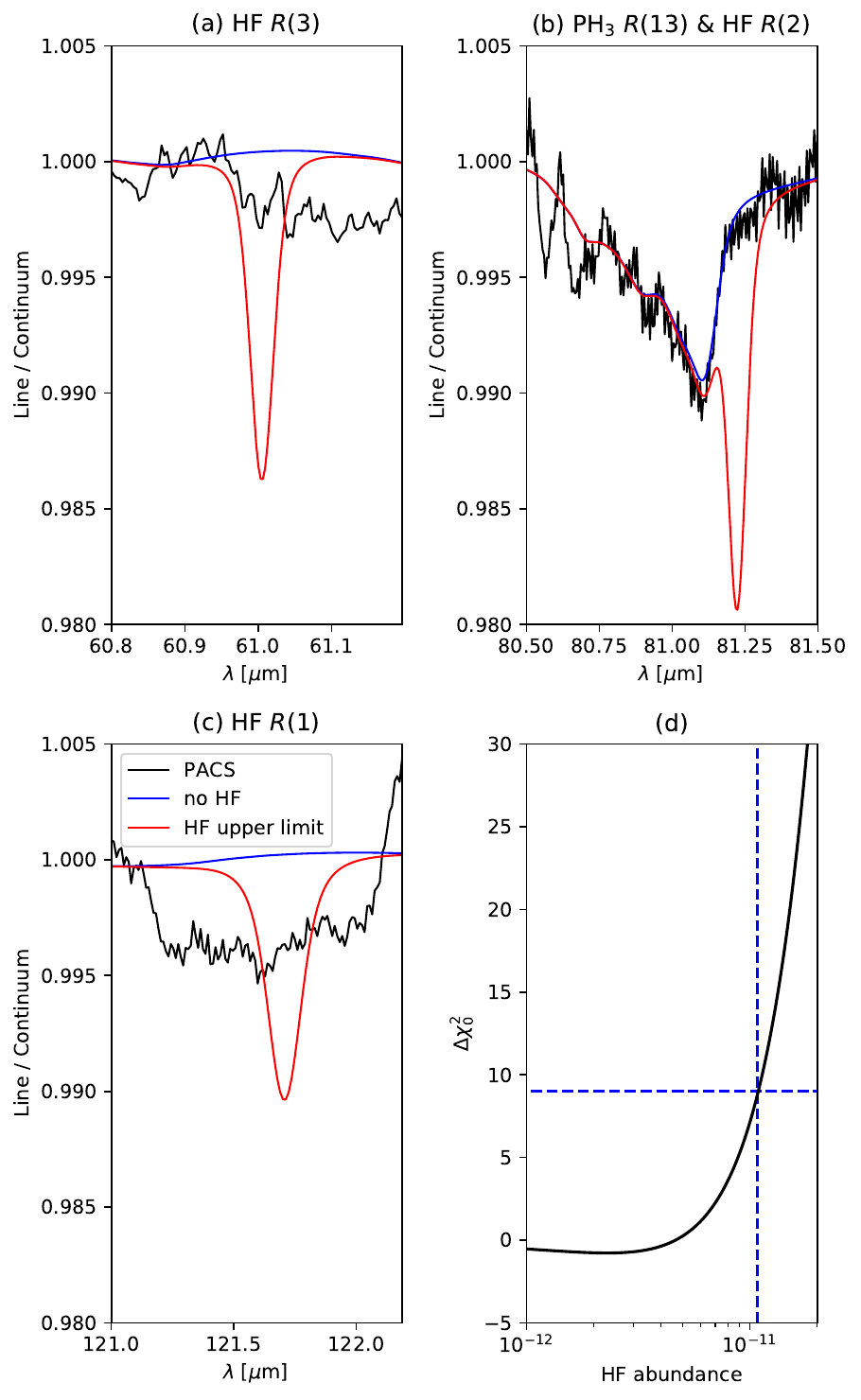}
	\caption{Results of the upper limit determination of Jupiter's atmospheric hydrogen fluoride (\ce{HF}) mole fraction using the PACS data and the a priori forward model (Sec. \ref{sec:model_setup}) including the derived ammonia and phosphine profiles (Subsec. \ref{subsec:nh3} and \ref{subsec:ph3}) and the inferred methane mole fraction (Subsec. \ref{subsec:ch4}). Panels (a) to (c) show the PACS data around the wavelengths of three \ce{HF} lines in the PACS spectral range. These plots also show forward model spectra including \ce{HF} with the inferred upper limit mole fraction and without \ce{HF} at all. Panel (d) shows the results of the least-squares comparison between model spectra assuming different \ce{HF} mole fractions and the PACS data using $\Delta\chi_0^2$ (see Equation \ref{eq:delta_chi-squared0}). The criterion for the upper limit ($\Delta\chi_0^2=9$) is shown with a horizontal blue dashed line and the mole fraction at which it is met is shown using a vertical blue dashed line. In total, 1000 values for the \ce{HF} mole fraction between 0 and $1\times 10^{-9}$ were used in the forward model to compare the resulting model spectra with the PACS data.}
	\label{fig:hf}
\end{figure}
\begin{figure}[!htbp]
	\centering
	\includegraphics[width=\hsize]{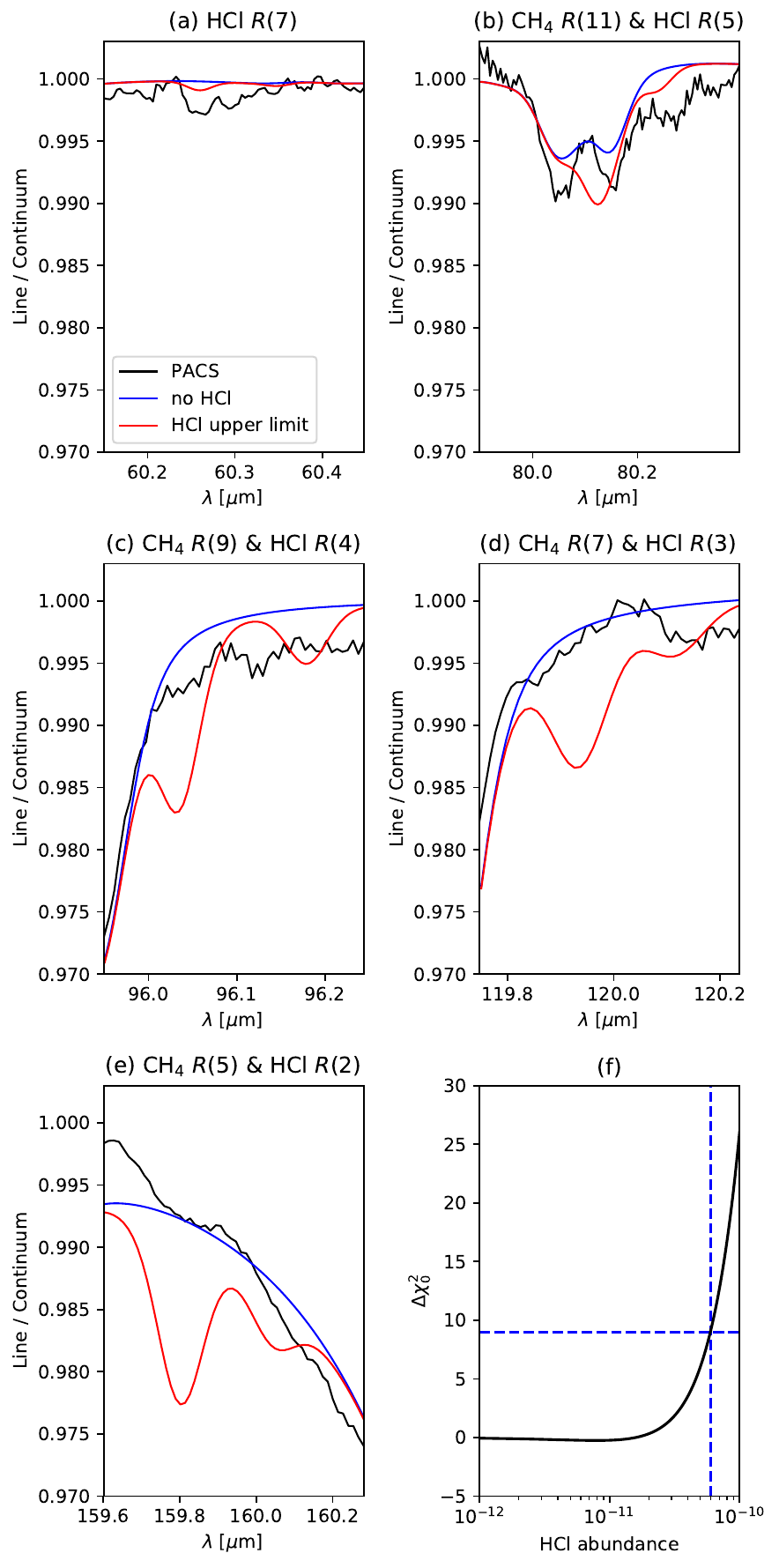}
	\caption{Same as Fig. \ref{fig:hf}, but for the results of the upper limit determination for the Jovian hydrogen chloride (\ce{HCl}) abundance. In total, 100 values for the \ce{HCl} mole fraction between 0 and $1\times 10^{-10}$ were used in the forward model to compare the resulting model spectra with the PACS data.}
	\label{fig:hcl_tropo}
\end{figure}
\begin{figure*}[htbp!]
    \centering
    \includegraphics[width=\hsize]{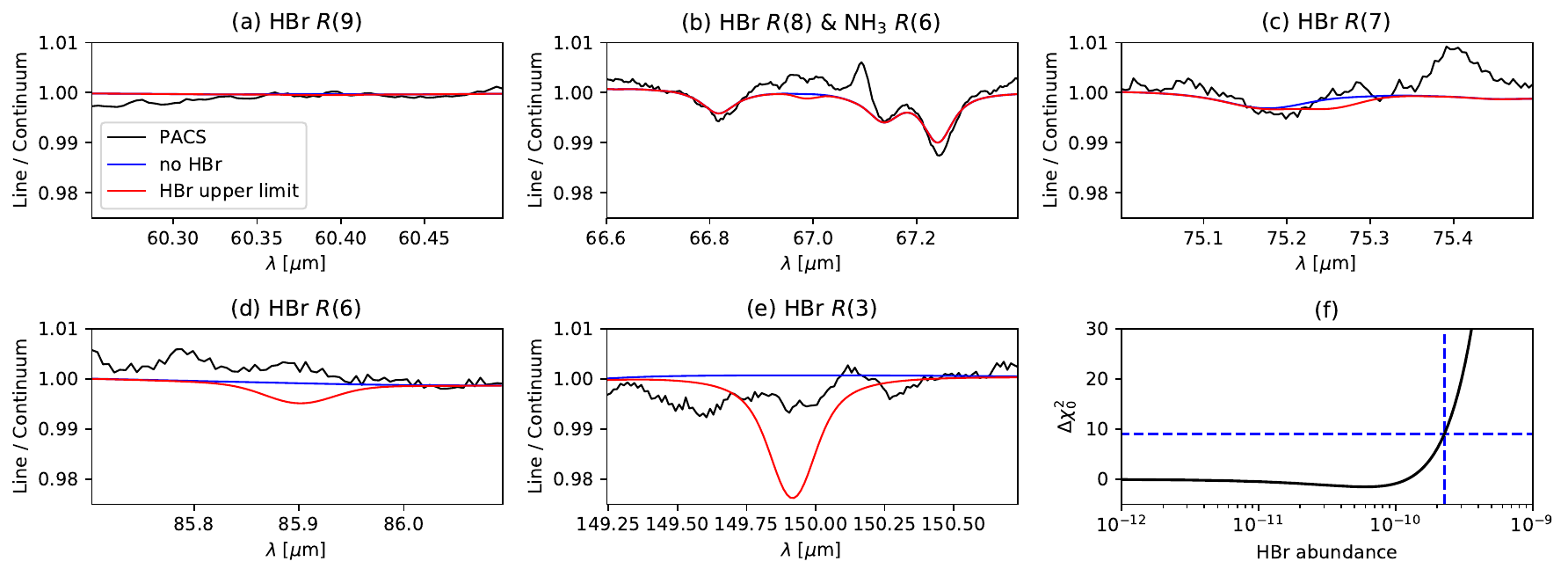}
    \caption{Same as Fig. \ref{fig:hf}, but for the results of the upper limit determination for the Jovian hydrogen bromide (\ce{HBr}) abundance. In total, 100 values for the \ce{HBr} mole fraction between 0 and $2\times 10^{-11}$ were used in the forward model to compare the resulting model spectra with the PACS data.}
    \label{fig:hbr}
\end{figure*}
\begin{figure*}[htbp!]
	\centering
	\includegraphics[width=\hsize]{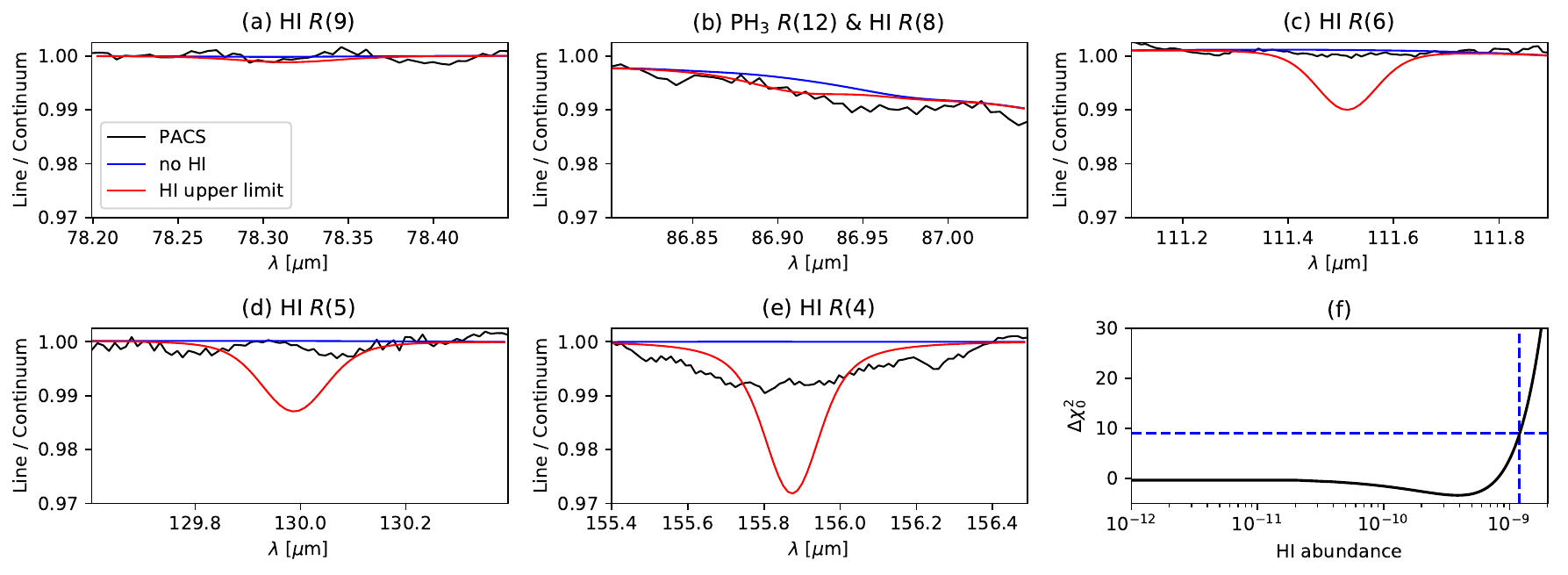}
	\caption{Same as Fig. \ref{fig:hf}, but for the results of the upper limit determination for the Jovian hydrogen iodide (\ce{HI}) abundance. In total, 100 values for the \ce{HI} mole fraction between 0 and $2\times 10^{-9}$ were used in the forward model to compare the resulting model spectra with the PACS data.}
	\label{fig:hi}
\end{figure*}
\begin{figure*}[htbp!]
	\centering
	\includegraphics[width=\hsize]{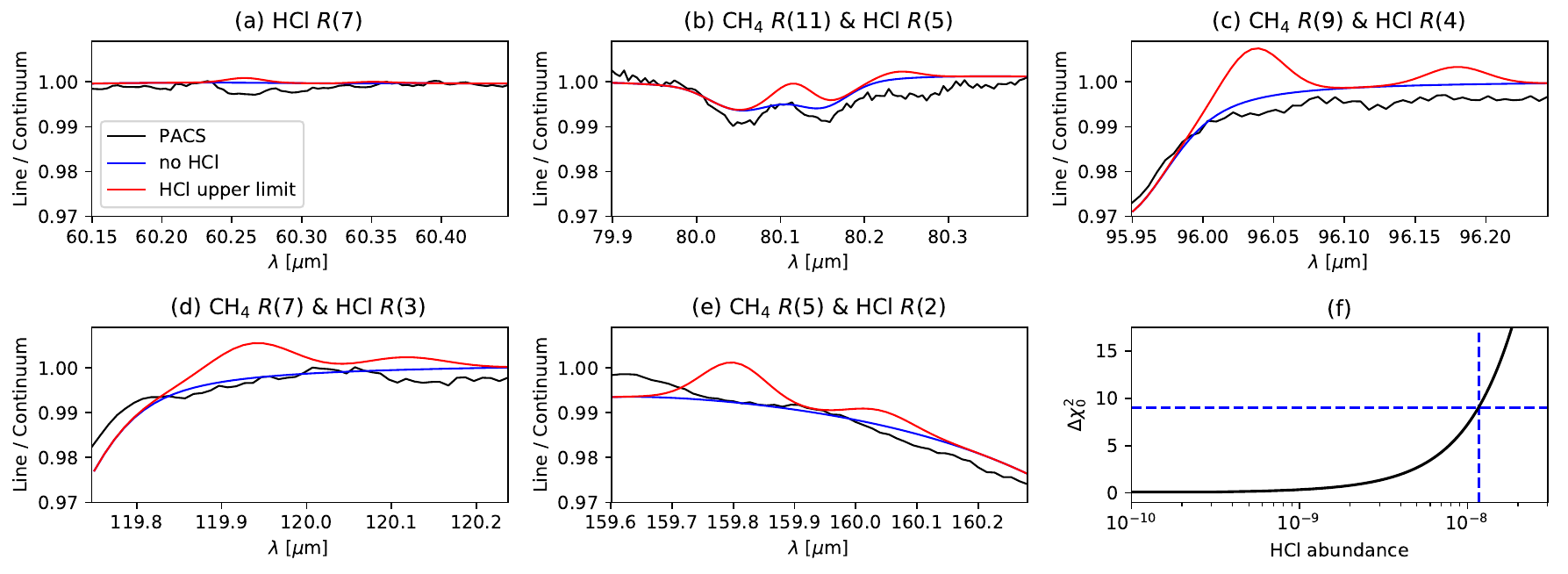}
	\caption{Same as Fig. \ref{fig:hf}, but for the results of the upper limit determination for the hydrogen chloride (\ce{HCl}) abundance in the Jovian stratosphere at pressures lower than $1\,$mbar. In total, 100 values for the stratospheric \ce{HCl} mole fraction between 0 and $3\times 10^{-8}$ were used in the forward model to compare the resulting model spectra with the PACS data.}
	\label{fig:hcl_strato}
\end{figure*}

\end{appendix}

\end{document}